\documentclass[]{aa}
\usepackage{graphicx}
\usepackage{rotating,subfigure,amssymb}
\usepackage{txfonts}
\usepackage{subfigure}
\def \xmm {\hbox{\it XMM-Newton }}
\def \chandra {\hbox{\it Chandra }}
\def\etal{et al.\ }
\def\betamod{$\beta$-model}
\def\nh{{N_{\rm H}}}\def\hzero{{H_{0}}}

\def\kT {{\rm k}T}

\def\Tx {T}
\def\Mv {M_{\rm 200}}
\def\keV {\rm keV}
\def\rs {r_{\rm s}}
\def \rv {r_{200}}
\def\rc {r_{\rm c}}
\def \ne {n_{\rm e}}
\def \hs {\rm h_{70}}

\def\msol{{M_{\odot}}}
\def \rhoDM {\rho_{\rm DM}}

\def \rhocz {\rho_{\rm c}(z)}

\begin{document}
     \title{\xmm observations of three poor clusters:  Similarity in
dark matter and entropy profiles  down to low mass} 

     \author{G. W. Pratt$^{1,2}$ \and M. Arnaud$^1$
     }
     \offprints{G. W. Pratt, \email{gwp@mpe.mpg.de}}

     \institute{$^1$ CEA/Saclay, Service d'Astrophysique,
                L'Orme des Merisiers, B\^{a}t. 709,
                91191 Gif-sur-Yvette Cedex, France\\
                $^2$ MPE Garching, Giessenbachstra{\ss}e, 85748
Garching, Germany} 
     \date{Received 16 June 2004 / Accepted 2 September 2004}

\abstract{We present an analysis of the mass and entropy profiles of
three poor galaxy clusters (A1991, A2717 and MKW9) observed with {\it
XMM-Newton}. The clusters have very similar temperatures ($\kT =
2.65$, 2.53 and 2.58 keV), and similar redshifts ($0.04 \lesssim z
\lesssim 0.06$). We trace the surface brightness, temperature, entropy
and integrated mass profiles with excellent precision up  to
$\sim500~h_{70}^{-1}$ kpc (A1991 and A2717) and $\sim350~h_{70}^{-1}$
kpc (MKW9).  This corresponds to $0.5(0.35)~\rv$, where $\rv$ is the
radius corresponding to a density contrast of 200 with respect to the
critical density at the cluster redshifts. None of the surface
brightness profiles is well fitted with a single $\beta$-model. Double
isothermal $\beta$-models provide reasonable fits, and in all cases
the value of the external $\beta$ parameter is consistent with the
value found for richer clusters. The temperature profiles have central
dips but are approximately flat at the exterior, up to the detection
limit. The integrated mass profiles are very similar in physical units
and are reasonably well fitted with the NFW mass model with
concentration parameters in the range $c_{200} =4-6$ and  $M_{200} =
1.2-1.6\times10^{14}~h_{70}^{-1}\msol$. A King model is inconsistent
with these mass data. The entropy profiles are very similar at large
scale, but there is some scatter in the very central region ($r
\lesssim 50$ kpc). However, none of the clusters has an isentropic
core.  

We then discuss the structural and scaling properties of cluster mass
and entropy profiles, including similar quality \xmm\ data on the
slightly cooler cluster A1983 ($\kT = 2.2$ keV), and on the massive
cluster A1413 ($\kT = 6.5$ keV). We find that the  mass profiles
scaled in units of $\Mv$ and $\rv$ nearly coincide, with $\lesssim 20$
per cent dispersion in the radial range $[0.05-0.5]~\rv$, where we
could compare the profiles without excessive extrapolation. We provide
a quantitative test of mass profile shapes by combining the
concentration parameters of these poor clusters with other values of
similar precision from the literature, and comparing with the
$c_{200}$--$M_{200}$ relation derived from numerical simulations for a
$\Lambda$CDM cosmology. The data  are fully consistent with the
predictions, taking into account the measurement errors and expected
intrinsic scatter, in the mass range $\Mv = [1.2 \times10^{14}-1.9
\times10^{15}]~h_{70}^{-1}\msol$. This excellent agreement with
theoretical predictions - a quasi universal cusped mass profile with
concentration parameters as expected - shows that the physics of the
dark matter collapse is basically understood. Scaling the entropy
profiles using the self-similar relation $S \propto T$, we find a
typical scatter of $\sim 30$ per cent in scaled entropy in the radial
range  $[0.05 - 0.5]~\rv$. The dispersion is reduced ($\sim 22$ per
cent) if we use the empirical relation $S \propto T^{0.65}$. The
scatter is nearly constant with radius, indicating a genuine
similarity in entropy profile shape. The averaged scaled profile is
well fitted by a power law for $0.05<r/\rv< 0.5$, with a slope
slightly lower than expected from pure shock heating ($\alpha =
0.94\pm0.14$), and a normalisation at $0.1~\rv$ consistent with
previous ROSAT/ASCA studies.  These precise XMM observations  confirm
that the entropy profiles of clusters are self-similar down to low
mass ($\kT\sim2~\keV$), but  that the entropy temperature relation is
shallower than in the purely gravitational model.  This
self-similarity of shape is a strong constraint, allowing us to rule
out simple pre-heating models. The gas history thus probably depends
not only on gravitational processes, but also on the interplay between
cooling and various galaxy feedback mechanisms.

\keywords{ Cosmology:
observations, Cosmology: dark matter, X-rays: galaxies: clusters,
galaxies: clusters: individual: \object{A 1991}, \object{A 2717},
\object{MKW 9}, Galaxies: clusters: Intergalactic medium } } 

     \authorrunning{G.W. Pratt \& M. Arnaud}
     \titlerunning{Three poor clusters observed with \xmm} 
\maketitle
%

\section{Introduction}

In galaxy clusters the hot, X-ray emitting gas of the intra-cluster
medium (ICM) lies trapped in the potential well of the dominant dark
matter component. If clusters were formed solely through gravitational
processes, the properties of different haloes (e.g., X-ray luminosity,
$L_X$, the gas mass $M_{\rm gas}$) would scale with  the mass $M$ (or
the global X-ray temperature $T_X$) and redshift, $z$, of the system,
such that $Q \propto A (z) T_X^{\alpha}$ (e.g., Bryan \&
Norman~\cite{bn}; Eke et al.~\cite{eke}). It has been known for some
time that while such relations do exist in observed clusters, their
actual scaling is subtly different from expected. 

With the advent of \xmm\ and {\it Chandra}, observations and numerical
simulations are on an equal footing, and so we can test the modelling
of the dark matter collapse and the consequent evolution of the ICM as
never before. We are beginning to see some evidence that numerical
simulations predict the correct shape for the dark matter distribution
from $\sim 0.01~\rv$ up to $\sim 0.7~\rv$ not only in massive systems
(e.g., Lewis, Buote \& Stocke~\cite{lewis}; Pratt \&
Arnaud~\cite{pa02}; Pointecouteau et al.~\cite{pointeco}), but also in
low mass clusters (Pratt \& Arnaud~\cite{pa03}), and that the shape is
independent of cluster temperature (e.g., Arnaud, Pratt \&
Pointecouteau~\cite{app}). Thus the underlying universality of the
dark matter distribution, indicated indirectly  by the similarity of
gas emission measure and temperature profiles  of relatively hot
systems (Markevitch \etal~\cite{mark98}; Neumann \&
Arnaud~\cite{na99}; Vikhlinin, Forman \& Jones~\cite{vikh99}; Irwin \&
Bregman~\cite{ib00}; De Grandi \& Molendi~\cite{demol02}; Arnaud,
Aghanim \& Neumann~\cite{aan02}) is starting  to be confirmed. 

At the same time, our knowledge of some aspects of the gas physics
seems wanting, as the departures from the expected scaling relations
attest. The current consensus is that there is some form of
non-gravitational process which affects the gas and thus modifies the
similarity. The non-gravitational processes have been historically
divided into either pre-heating, where the gas has been heated before
being accreted into the potential well, by early supernovae and/or AGN
activity (e.g. Kaiser~\cite{kaiser91}; Evrard \&
Henry~\cite{evrard91}; Valageas \& Silk~\cite{valageas}), internal
heating after accretion  (e.g. Metzler \& Evrard~\cite{me94}), and
cooling (e.g. Pearce \etal ~\cite{pea00}). A great deal of theoretical
effort, including the use of numerical simulations, has been put into
determining how these processes affect the scaling properties of the
ICM (e.g. see Borgani \etal~\cite{borg04} for a review). 

The preferred quantity in which to cast both observations and theory
is the entropy, (see e.g., Bower~\cite{bow97}). The entropy reflects
the accretion history of the gas, but at the same time is likely to
preserve the imprint of any other physical process. It has become
traditional to define the `entropy' as $S = \kT/\ne^{2/3}$,  which is
related to the true thermodynamic entropy via a logarithm and an
additive constant. Pioneering work on cluster entropy profiles can be
found in David, Jones \& Forman~(\cite{djf}).  

In the self-similar framework described above, the entropy scales with
the temperature simply as $S \propto T$. Ponman, Cannon \&
Navarro~(\cite{pcn99}) used {\it ROSAT\/} observations to suggest that
the entropy measured at $0.1~\rv$ follows the standard self-similar
scaling at high temperatures but tends towards a limiting value in the
coolest systems. This sparked an avalanche of work on
non-gravitational heating (but see also Bryan~\cite{bry00}; Dav\'e et
al.~\cite{dav02}), with a particular focus on the pre-heating scenario
(e.g. Fujita \& Takahara~\cite{fuj00}; Bialek \etal~\cite{bia01};
Tozzi \& Norman~\cite{tn01}; Babul \etal~\cite{bab02}; Borgani
\etal~\cite{borg02}); but no definitive picture emerged concerning the
level, actual timescale, or astrophysical source of the extra energy
injection (Wu, Fabian \& Nulsen~\cite{wfn98};
Loewenstein~\cite{loe00}; Kravtsov \& Yepes~\cite{kra00}; Bower
\etal~\cite{bow01}; Brighenti \& Mathews~\cite{bri01}; Yamada \&
Fujita~\cite{yam01};  Nath \& Roychowdhury~\cite{nat02}).   However,
more recent work has shown that there is no limiting value to the
entropy, and that in fact it scales as $S \propto T^{0.65}$ (Ponman,
Sanderson \& Finoguenov~\cite{psf03}). At the same time, better
spatially resolved \xmm and \chandra observations have shown that the
simple prescriptions for gas pre-heating, intended to match the
limiting value on the entropy, were predicting large isentropic cores
in groups when in fact there are none (Pratt \& Arnaud~\cite{pa03};
Sun et al.~\cite{sun}; Rasmussen \& Ponman~\cite{rp}; Khosroshahi,
Jones \& Ponman~\cite{khos}).  

\begin{table}
\begin{minipage}{\columnwidth}
\caption{{\footnotesize Journal of observations.}}\label{tab:obs}
\centering
\begin{tabular}{l l l l l l l}
\hline
\hline
\multicolumn{1}{l}{Cluster} & \multicolumn{1}{l}{z} &
\multicolumn{1}{l}{Rev.} & \multicolumn{1}{l}{Mode$^a$} &
\multicolumn{3}{c}{$t_{\rm exp}^b$} \\ 
\multicolumn{1}{l}{ } & \multicolumn{1}{l}{} & \multicolumn{1}{l}{} &
\multicolumn{1}{l}{ } & \multicolumn{1}{l}{MOS1} &
\multicolumn{1}{l}{MOS2} & \multicolumn{1}{l}{pn} \\ 
\hline
A1991 & 0.0586 &584 &  FF & 28975 & 28878 & 19348 \\
A2717 & 0.0498 &558 & FF & 51682 & 51647 & 44284 \\
MKW9  & 0.0382 &311 & EFF & 29890 & 29716 & 20720 \\
\hline
\end{tabular}
Notes: $^a$ EPN observation mode: FF = Full Frame, EFF = Extended Full Frame; 
$^b$  Exposure time after flare cleaning
\end{minipage}
\end{table}

The focus has now started to shift from simple pre-heating to more
advanced models which reflect the energy balance between heating and
cooling, and the effect of heating on the accretion of the gas
(e.g. Menci \& Cavaliere~\cite{men00}; Muawong \etal~\cite{mua02};
Finoguenov \etal~\cite{fin03}; Xue \& Wu~\cite{xw03};
Valdarnini~\cite{vald03}; Kay, Thomas \& Theuns~\cite{kay03}; Voit et
al.~\cite{voit02}, \cite{voit03}; Voit \& Ponman~\cite{vp03};
Kay~\cite{kay04}; Borgani \etal~\cite{borg04}). These models have need
of stringent constraints from observations, which \xmm\ and \chandra
are starting to provide.  

As shown by Voit et al.~(\cite{voit02},~\cite{voit03}), the two
fundamental quantities which define the X-ray properties of a relaxed
cluster are the entropy profile of the gas $S(r)$ and the shape of the
potential well in which it lies. $M(r)$ and $S(r)$ reflect
respectively the physics of the gravitational collapse and
thermodynamic history of the gas. Low mass clusters are particularly
interesting targets for the study of the entropy because in these
systems gravitational and non-gravitational processes affect the ICM
in roughly equal proportions. In this paper, we present the results
from high quality \xmm\ observations of three cool clusters, A1991,
A2717 and MKW9, focussing on the gas  density and temperature
profiles (Sect.~\ref{sec:gasdens} and
Sect.~\ref{sec:temp})\footnote{We  will discuss the abundance
distributions in a companion paper (Pratt \& Arnaud, in prep).}. We
derive the radial distributions of these quantities with good
precision between $\sim 0.01$ and $0.5$ of the virial radius ($\rv$),
enabling us to calculate the total mass profiles
(Sect.~\ref{sec:mass}) and entropy profiles (Sect.~\ref{sec:entropy})
out to the same distance with unprecedented accuracy. To investigate
the scaling and structural properties of cluster mass and entropy
profiles, we combine these data  with previously published results for
the cooler group A1983 and the hot cluster A1413. We will see that the
dispersion in the scaled mass and entropy profiles is remarkably
small. We discuss the implications for our understanding of the dark
matter collapse and the gas specific physics in Sect.~\ref{sec:dis}.  

Except where otherwise noted, the following cosmology is used
throughout: $\hzero = 70$ km s$^{-1}$ Mpc$^{-1}$, $\Omega_m = 0.3$ and
$\Omega_{\Lambda} = 0.7$, which is denoted $\Lambda$CDMH70. Where we
want to examine the dependence of scaling on cosmology, we will also
use the SCDMH50 cosmology: $H_0 = 50$ km s$^{-1}$ Mpc$^{-1}$,
$\Omega_{\rm m} = 1.0$ and $\Omega_{\Lambda} = 0.0$. 
\begin{figure*}
\begin{centering}
\includegraphics[scale=0.45,angle=0,keepaspectratio]{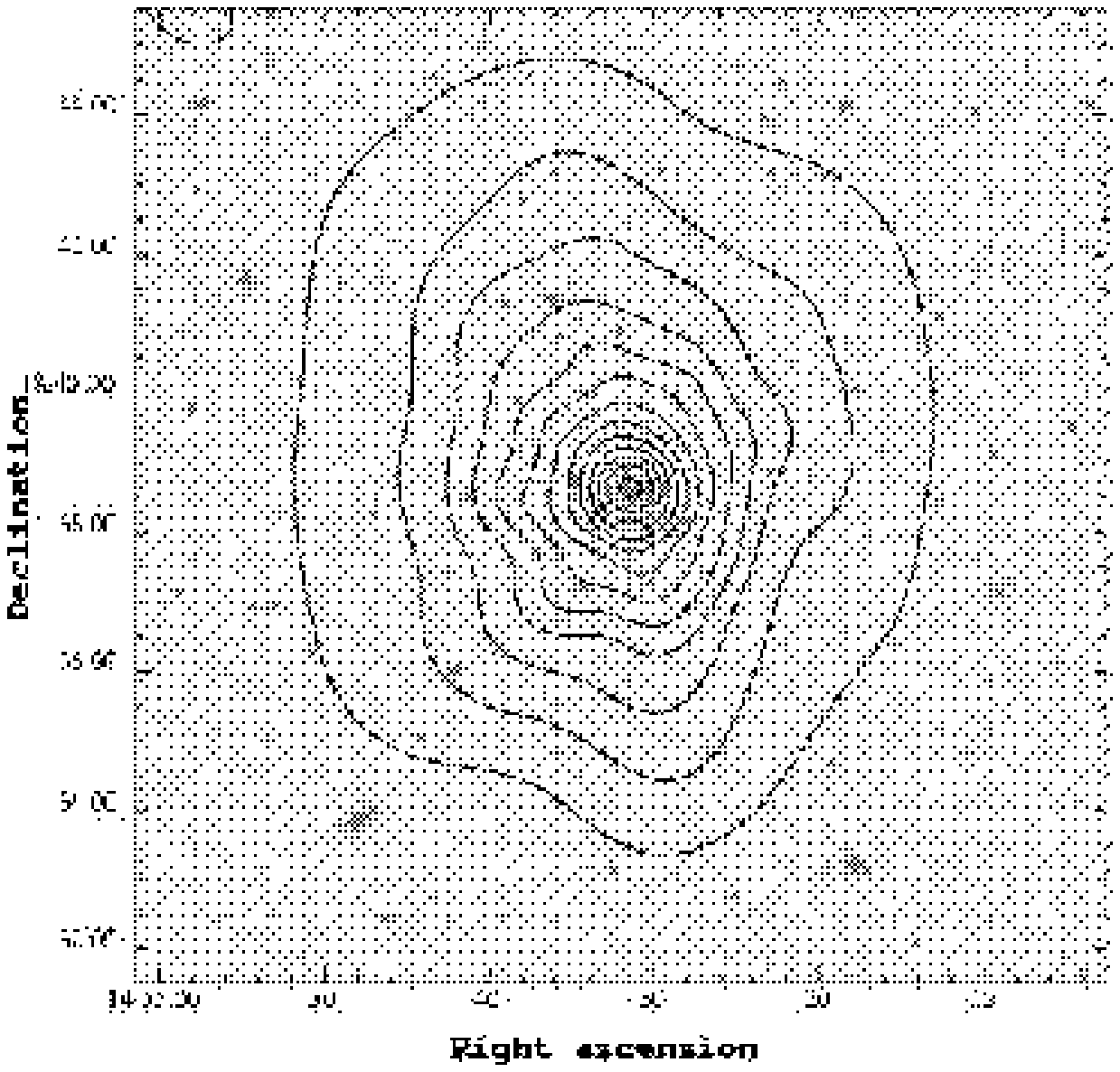}
\includegraphics[scale=0.45,angle=0,keepaspectratio]{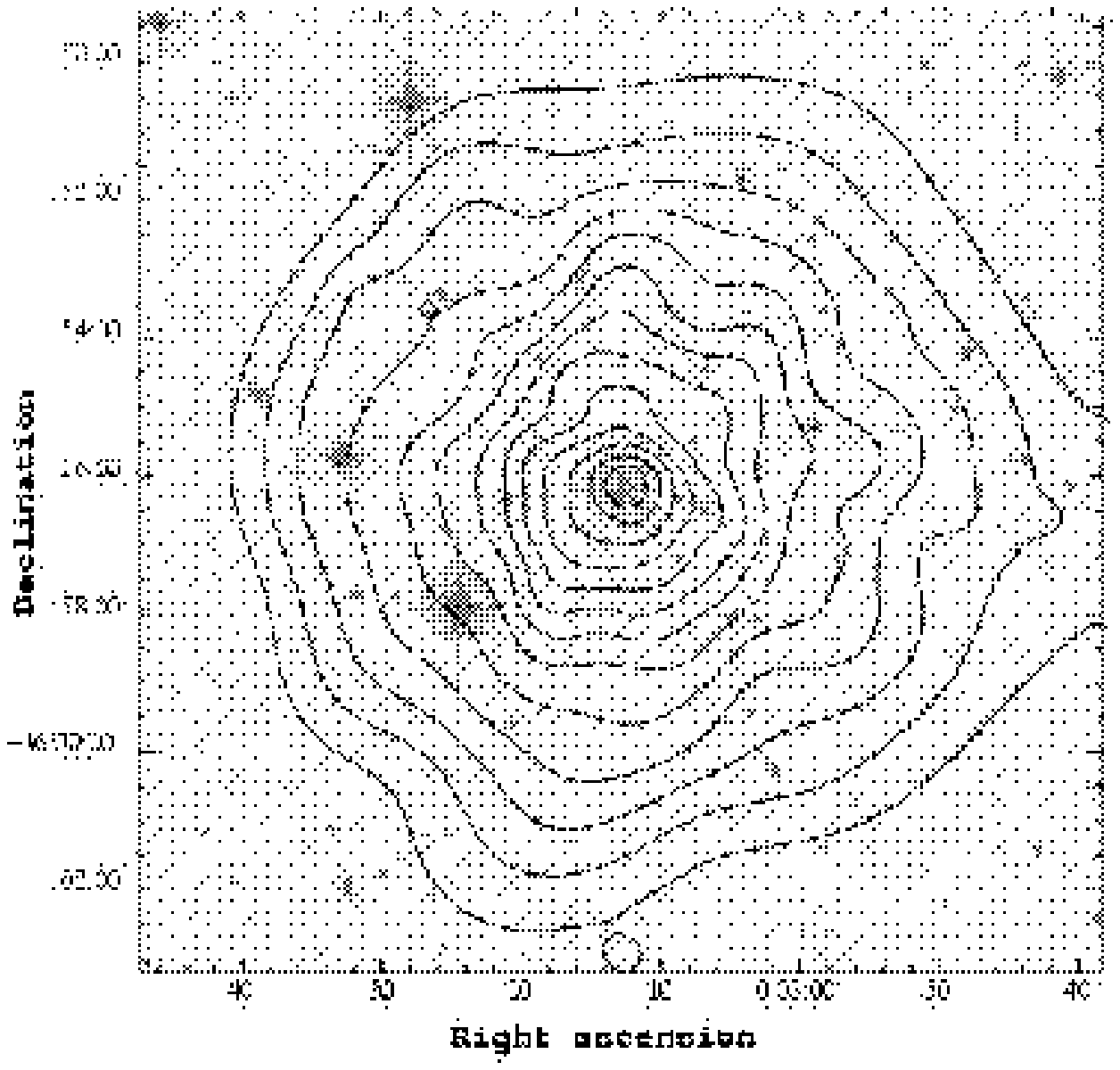}
\includegraphics[scale=0.45,angle=0,keepaspectratio]{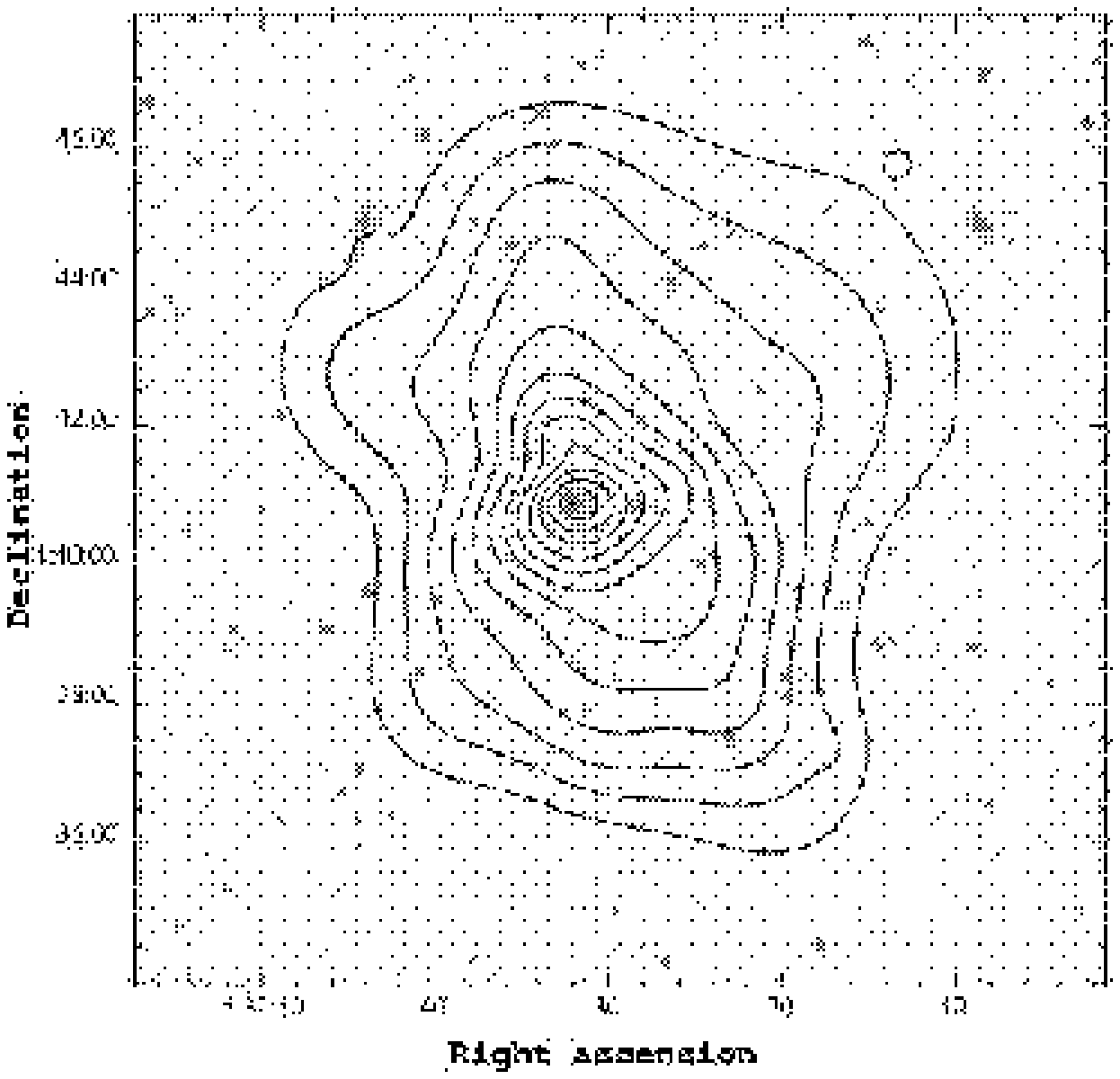}
\caption{{\footnotesize XMM/DSS overlay images of each cluster. X-ray
contours come from the adaptively smoothed, non-background subtracted,
[0.3-0.9] keV EMOS+EPN image, and are logarithmically
spaced.}}\label{fig:optxim} 
\end{centering}
\end{figure*}

\section{Observations and data reduction}

We show the observation details in Table~\ref{tab:obs}. For the
following analysis we use PATTERNs 1-12 from EMOS event lists, and
PATTERN 0 from the EPN event list. Each observation was cleaned for
soft proton flares according to the method described in Pratt \&
Arnaud~(\cite{pa02}), using an initial pass in 100s bins in a
high-energy band (10.-12. keV for EMOS; 12.-14. keV for EPN), with a
second pass in 10s bins in a broad band (0.3-10. keV). To correct for
vignetting, the photon-weighting method of Arnaud et
al.~(\cite{arnaudetal01a}), as implemented in the SAS task {\tt
evigweight}, was applied to each event file. 

We use the dedicated blank-sky event lists accumulated by Read \&
Ponman~(\cite{read}) as background files, applying the same PATTERN
selection, flare rejection criteria and vignetting correction as
described above. Background subtraction of output products was
undertaken as described in Arnaud \etal~(\cite{aml02}).

\section{Gas density distributions}
\label{sec:gasdens}
\subsection{Morphologies}

The X-ray morphology can give interesting qualitative (and
quantitative, see e.g., Buote \& Tsai~\cite{bt96}) insights into the
dynamical status of a given cluster. The positions of bright
serendipitous point sources were obtained from the combined EPIC
source list in the pipeline products. We then used the CIAO utility
{\tt dmfilth} to excise these sources from the images and refill the
holes with Poisson noise, and then we used {\tt csmooth} to adaptively
smooth the images. In Figure~\ref{fig:optxim}, we show an XMM/DSS
overlay of each cluster, where the contours come from the adaptively
smoothed [0.3-0.9] keV EMOS+EPN image. 

\object{A1991} and \object{A2717} exhibit symmetric X-ray isophotes,
suggesting that they are relatively relaxed. The X-ray emission of
A2717 is centred on its central D galaxy [PL95] ACO 2717 BCG, the
Brightest Cluster Galaxy as defined by Postman \& Laurer~(\cite{PL}),
which has coordinates $00^{\rm h} 03^{m} 12\fs89$, $-35^{\circ}
56\arcmin 12\farcs3$. Note that the SIMBAD coordinates for this
cluster are currently offset by $6\arcmin$. The X-ray emission of
A1991 is slightly offset from the optical position of the central D
galaxy [WCB96] ACO 1991 A (aka NGC 5778)\footnote{In fact, the raw
image seems to show a surface brightness edge $\sim 10\arcsec$ from
the X-ray centroid. These \xmm\ data do not have sufficient spatial
resolution to resolve the feature, but the  {\it Chandra} observation
(Sharma \etal \cite{sharma}) does show substructures at that scale.},
coordinates $14^{\rm h} 54^{m} 31\fs54$, $+18^{\circ} 38\arcmin
31\farcs1$.   

The X-ray emission is of \object{MKW9} is centred on the central
galaxy UGC 9886, coordinates $15^{\rm h} 32^{m} 32\fs16$, $+04^{\circ}
40\arcmin 51\farcs0$, but in contrast to the other clusters, the X-ray
morphology is not symmetric at all. This is especially evident at
large scale, where the isophotes have a pronounced ellipticity in the
NE-SW direction. At small scale the isophotes are preferentially
orientated in the E-W direction.  

\begin{figure*}
\begin{centering}
\includegraphics[scale=1.0,angle=0,keepaspectratio]{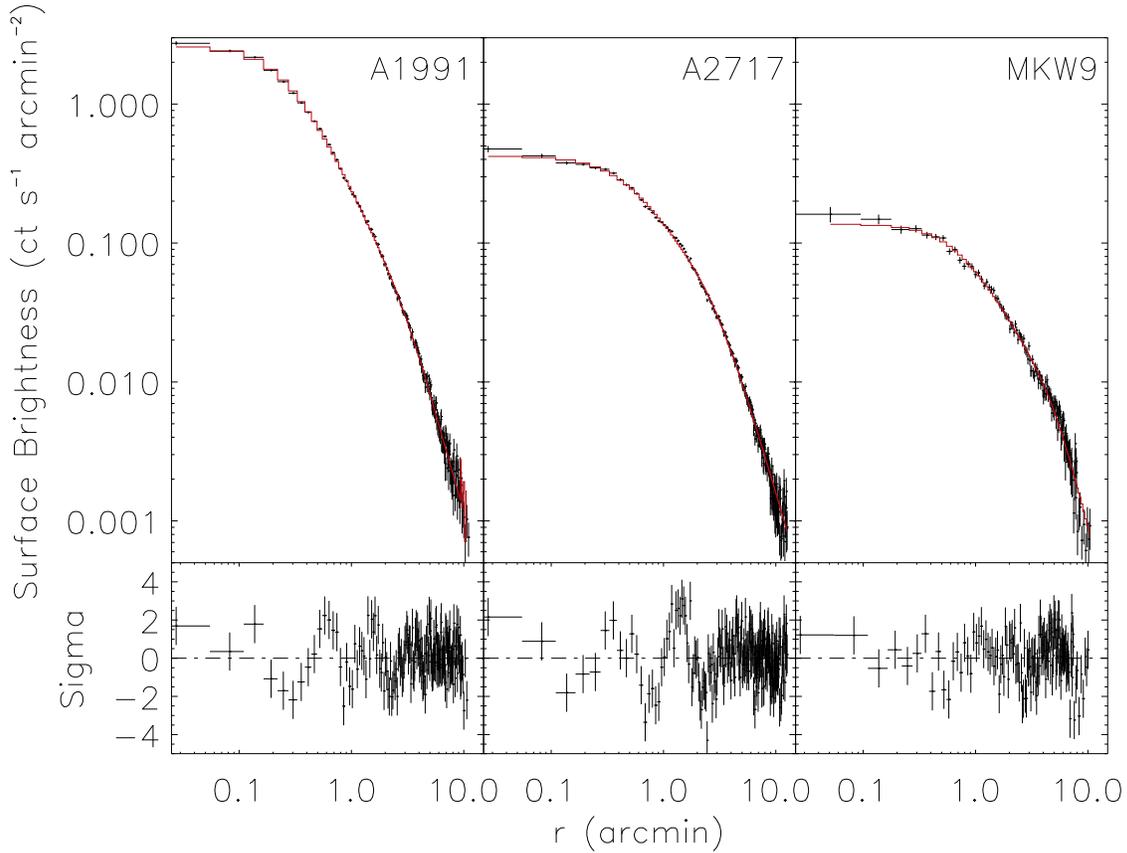}
\caption{{\footnotesize The combined EPIC surface brightness profile
of each cluster in the [0.3-3.0] keV band, with the [0.9-1.2] keV band
excluded and corrected for the dependence of the emissivity,
$\Lambda$, on radial temperature and abundance variations. Each
profile is background subtracted and corrected for vigentting. The
solid line in each case is the best-fitting BB model convolved with
the \xmm\ PSF; best-fit parameter values are given in
Table~\ref{tab:sbfit}.}}\label{fig:sbplot} 
\end{centering}
\end{figure*}

\subsection{Surface brightness profiles}

In poor clusters, the emissivity of the hot gas, $\Lambda$, depends
sensitively on the abundance and temperature. As will be seen below,
radial gradients in these quantities exist in all three clusters, and
so in deriving gas density profiles from the surface brightness
distribution, such variations should be taken into account.  

For each cluster, after masking of point sources, the surface
brightness profile  was extracted from each event list in the [0.3 -
3.0] keV band, with the [0.9-1.2] keV band excluded\footnote{This
minimises the contribution from the FeL blend, the feature which is
the most sensitive to abundance and temperature variations.}. Events
were binned directly from the event lists in circular annuli centred
at the X-ray emission peak. Background subtraction was undertaken
separately for each camera; the resulting EMOS and EPN data were
coadded and then the total profile binned such that a $S/N$ ratio of
at least $3\sigma$ was reached. 

The surface brightness profile was then corrected for emissivity
variations as follows. The temperature and abundance values were
interpolated to each radius in the surface brightness profile by
fitting the temperature and abundance profiles with functional forms
(respectively: an empirical function described in Eq.~\ref{eqn:allen},
and a lognorm function). The emissivity at each radius, $\Lambda
(\theta)$, was then estimated using an absorbed, redshifted MEKAL
model convolved with the instrument response. The surface brightness
profile was then divided by $\Lambda (\theta)$ normalised to its value
at large radii.  

\begin{table}
\begin{minipage}{\columnwidth}
\centering
\center
\caption{{\footnotesize Results of the double $\beta$ (BB)  model
analytical fits to the gas surface brightness profile, errors are $90$
per cent confidence.}} 
\begin{tabular}{ l l l l}
\hline
\hline
Parameter  & A1991  & A2717& MKW9 \\
\hline
$n_{\rm e,0} ({\rm h_{70}^{1/2}~cm^{-3}})$ & $5.61 \times 10^{-2}$ &
$1.23\times 10^{-2}$ & $7.86 \times 10^{-3}$\\ 
$r_{\rm c}$ & $1\farcm42^{+0.21}_{-0.19}$ &
$1\farcm95^{+0.15}_{-0.13}$ & $3\farcm36^{+1.08}_{-0.66}$ \\ 
$\beta$     & $0.65^{+0.02}_{-0.02}$  & $0.63^{+0.02}_{-0.01}$  &
$0.7^{+5.0}_{-0.1}$  \\ 
$R_{\rm cut}$   & $2\farcm08^{+0.23}_{-0.22}$ &
$2\farcm08^{+0.19}_{-0.17}$ &  $3\farcm40^{+0.60}_{-0.30}$ \\ 
$r_{\rm c, in}$ & $0\farcm16^{+0.01}_{-0.01}$ &
$0\farcm31^{+0.04}_{-0.04}$ &  $0\farcm35^{+0.05}_{-0.03}$ \\  
$\chi^2$/d.o.f  & 197.0/160 & 339.3/206 & 181.3/133 \\
\hline
\end{tabular}
\label{tab:sbfit}
\end{minipage}
\end{table} 

\subsection{Gas density profile modelling}
\label{sec:gasden}

The corrected surface brightness profile of each cluster (shown in
Fig.~\ref{fig:sbplot}) is directly proportional to the emission
measure profile, $EM(r)$, and can thus be fitted with various
parametric models for the gas density profile, $n_e (r)$. These models
were convolved with the \xmm\ PSF
(Ghizzardi\etal~\cite{ghizzardi01},~\cite{ghizzardi02}) and binned
into the same bins as the profile under consideration. 

In none of these clusters did we find that a standard \betamod\ was a
good description of the entire profile. In all cases the fit improves
and the reduced $\chi^2$ converges to 1.0 as the central regions are
progressively excluded, which indicates that the outer regions are in
fact well described with this model. We thus fitted the surface
brightness profiles with the double isothermal \betamod\ described in
Pratt \& Arnaud~(\cite{pa02}, their BB model). The fit results are
shown in Table~\ref{tab:sbfit}. Note that the profile obtained beyond
$R_{\rm cut}$ in the BB model is the classic \betamod, and that in all
cases the external $\beta$ value is consistent with the mean $\beta$
value for hotter clusters (Neumann \& Arnaud~\cite{na99}). 

\begin{figure*}
\begin{centering}
\includegraphics[scale=0.25,angle=0,keepaspectratio]{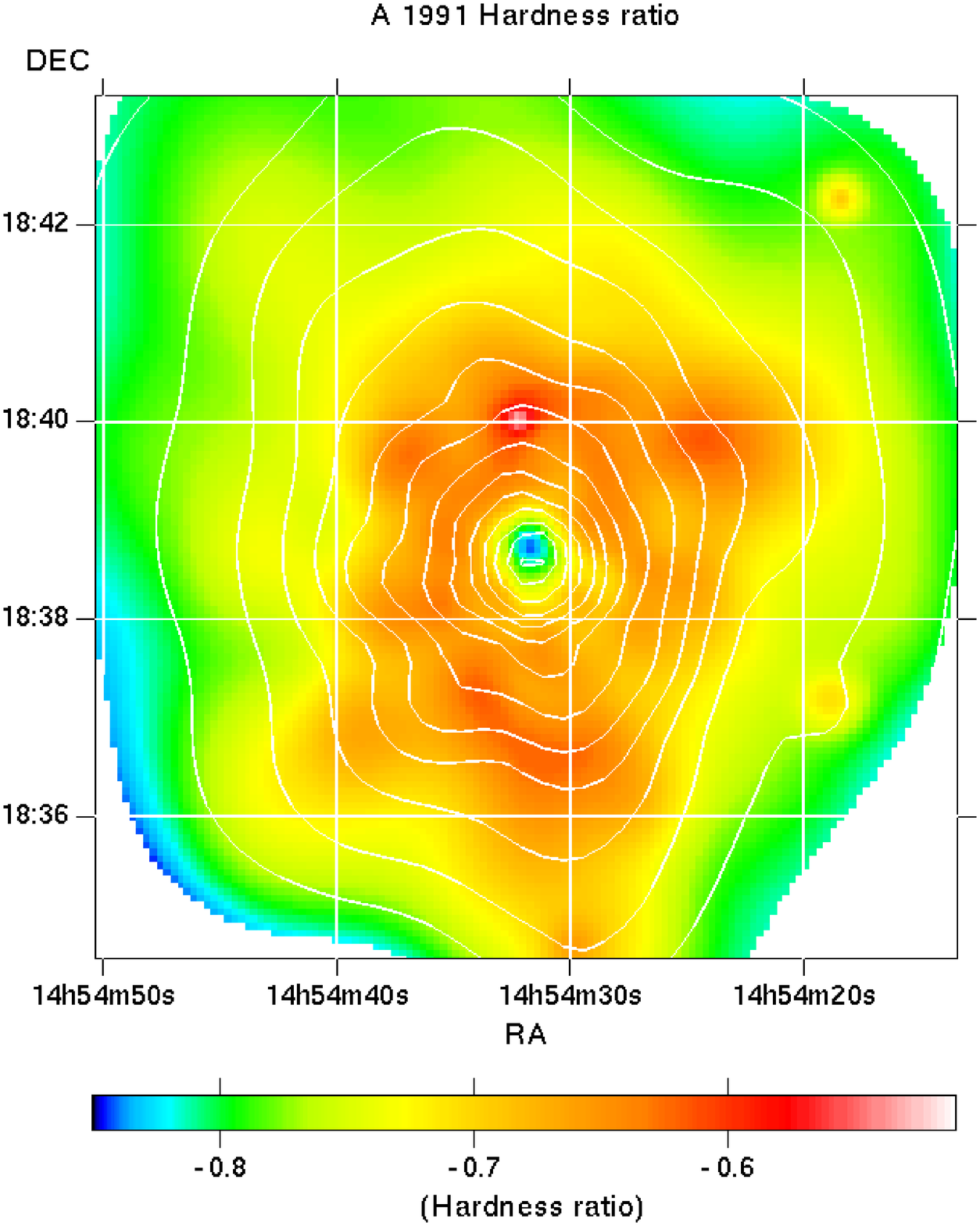}
\includegraphics[scale=0.25,angle=0,keepaspectratio]{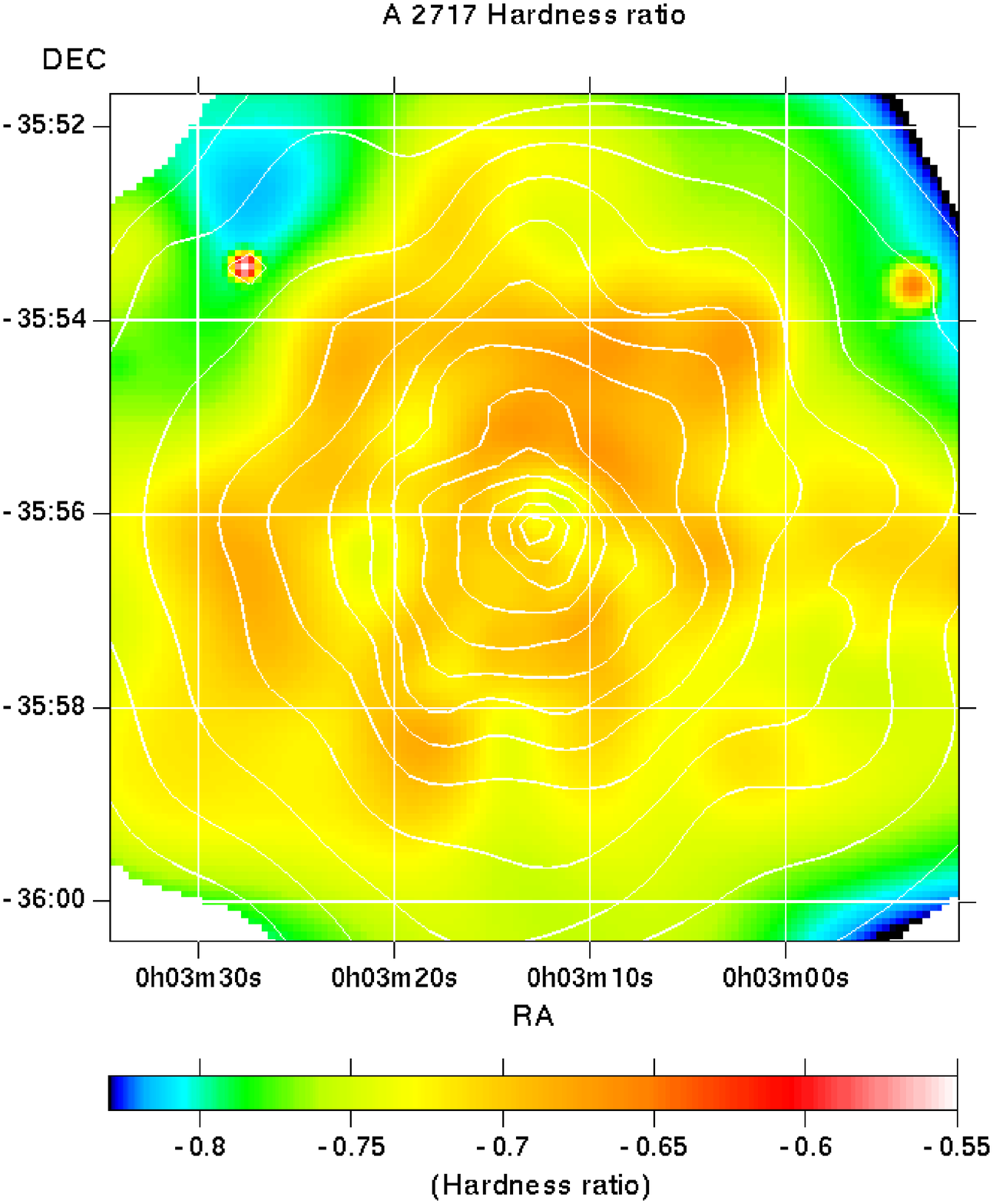}
\includegraphics[scale=0.25,angle=0,keepaspectratio]{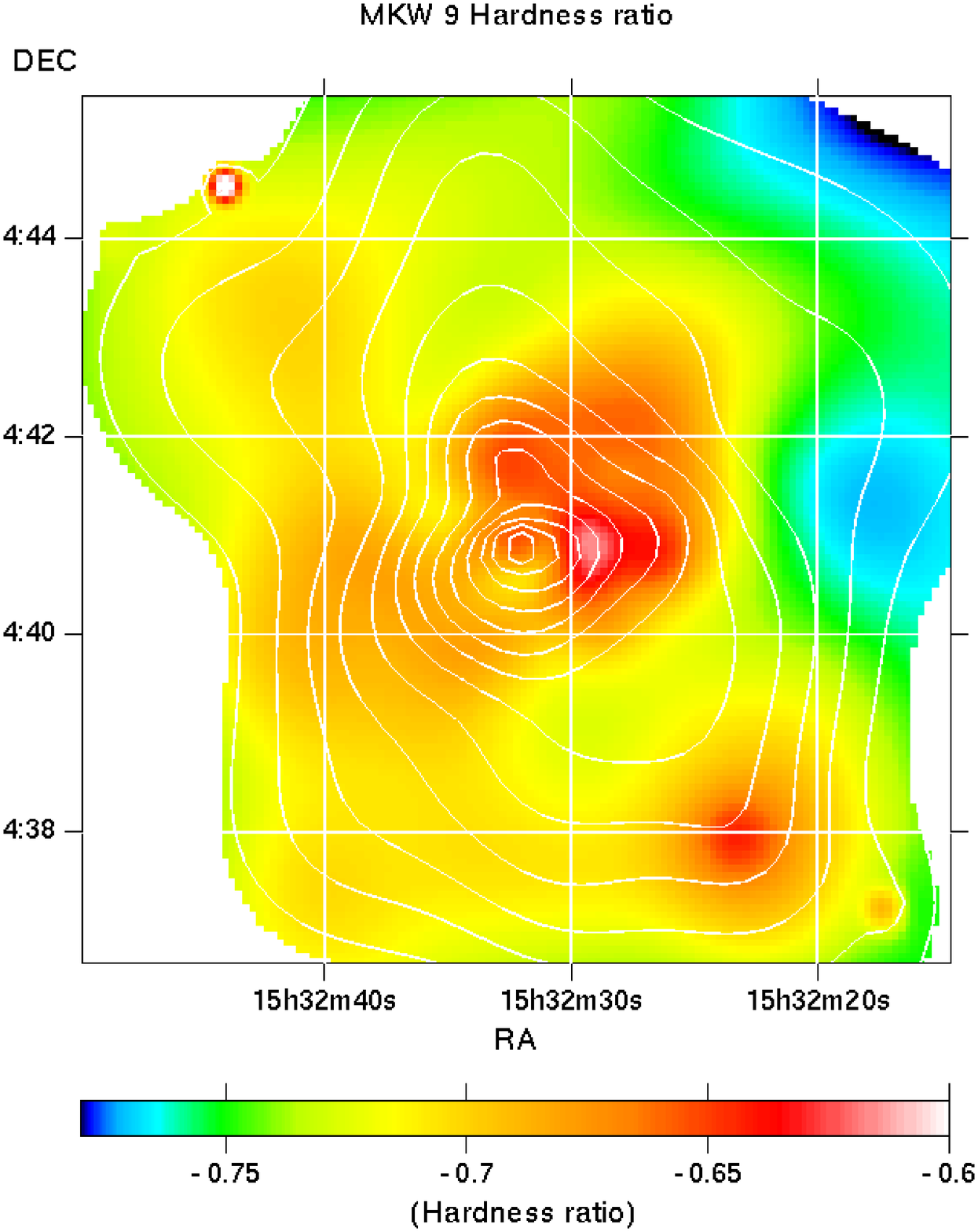}
\caption{{\footnotesize The hardness ratio (HR) image of each cluster
is overlaid with the same contour as in Fig.~\ref{fig:optxim}. These
images should be viewed bearing in mind that the difference in local
background and blank-sky background at low energy has {\it not\/} been
corrected for. {\it See the electronic edition of the Journal for a
colour version of this figure.}}}\label{fig:hrim}   
\end{centering}
\end{figure*}

\section{Temperature distributions}
\label{sec:temp}

\subsection{Hardness ratio images}

The hardness ratio (HR) of a source is an indirect measure of its
temperature, and taking advantage of the large bandpass of {\it
XMM-Newton}, we can use the information contained in images in
different energy bands to get a good idea of the projected temperature
structure of a given cluster. The method we use follows closely that
described in Pratt \& Arnaud~(\cite{pa03}), except that here we take
advantage of the CIAO utility {\tt dmfilth} to excise and refill with
Poisson noise the most obvious point sources, and then we use {\tt
csmooth} to adaptively smooth the images.  

Source and background images were extracted in the [0.3-0.9] keV and
[2.5-5.0] keV bands, chosen to avoid complications caused by the
coupling of temperature and abundance at low temperatures due to
strong line emission. This is especially important in these low
temperature clusters. We used the total EMOS+EPN [2.5-5.0] keV image
to define a smoothing template, and subsequently applied this template
to each source, background and exposure map image in turn. We tried
various scales and found that a minimum smoothing scale of $2.5\sigma$
and a maximum smoothing scale of $4\sigma$ was a good compromise
between under- and over-smoothing. The backgrounds were normalised
based on the effective exposure times and subtracted from each image,
then the images were divided by their exposure maps. The hardness
ratio image of each cluster was then calculated using HR = (image
[2.5-5.0] - image [0.3-0.9])/(image [2.5-5.0] + image [0.3-0.9]). 

We have not corrected for the difference between the local cluster
backgrounds and the blank-sky background at low energies, because the
effect  --- being energy-dependent --- is rather difficult to correct
for. This means in particular that i) the absolute HR values shown in
Fig.~\ref{fig:hrim} cannot reliably be converted into temperatures,
and ii) the decline in HR towards the outer regions is an
artifact. The latter point is explicitly confirmed  in
Sect.~\ref{sec:tprof} below. Despite these caveats, the HR images give
an excellent idea of the temperature {\it structure}, which is the
most interesting quantity for investigating the dynamical state of a
given cluster.  The HR images of A1991 and A2717 appear relatively
symmetric, with a clear temperature drop in the centre of A1991. On
the other hand, the HR image of MKW9 shows sub-structure, suggesting,
as does the raw image, that it may not be fully relaxed. This is
further discussed in Sect.~\ref{sec:disdm}.  


\subsection{Radial temperature profiles}
\label{sec:tprof}
\begin{figure}
\begin{centering}
\includegraphics[scale=1.,angle=0,keepaspectratio,width=\columnwidth]{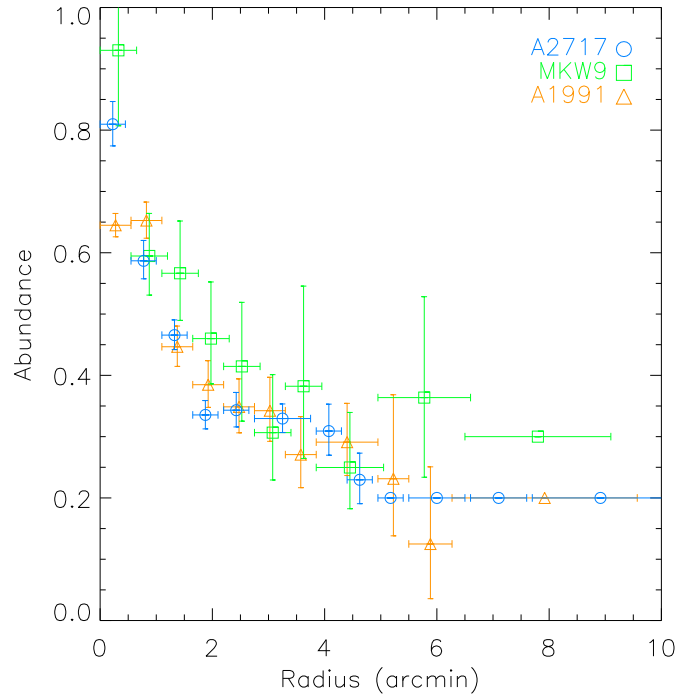}
\caption{{\footnotesize Cluster abundance profiles, with abundances
measured relative to Anders \& Grevesse~(\cite{agr}). Annuli where the
abundance was fixed (due to insufficient signal for an accurate
abundance measurement) have no error bars. {\it See the electronic
edition of the Journal for a colour version of this
figure.}}}\label{fig:abund}   
\end{centering}
\end{figure}

\subsubsection{ Annular spectral analysis}
Spectra were extracted in circular annuli centred on the X-ray
emission peak of each cluster. We optimised the annuli by imposing a
$5\sigma$ detection, after background subtraction, in the [2.0 - 5.0]
keV band; this was possible for all but the most external annuli, the
temperature determinations of which should be treated with
caution. Regardless of the detection significance, we additionally set
a lower annulus width of $0\farcm5$, to minimise PSF effects (i.e.,the
annular widths were greater than or equal to the diameter enclosing
$\sim 70$ per cent of the energy  for the on-axis PSF). 

The EMOS and EPN spectra of each cluster were fitted with an absorbed
{\sc MEKAL} model. The absorption was fixed at the galactic $\nh$
value in the direction of each cluster\footnote{If the absorption is
left as a free parameter, the absorption values for A2717 and MKW9
compare favourably with those measured in the HI survey by Dickey \&
Lockman~(\cite{dl90}), but that for A1991 is significantly
higher. However, since leaving the absorption free has no discernable
effect on the measured temperatures and abundances of any of these
clusters, we have chosen to fix the $\nh$ to the galactic value.}, as
measured in the HI survey by Dickey \& Lockman~(\cite{dl90}). The free
parameters of each fit are thus the temperatures and abundances
(measured relative to Anders \& Grevesse~\cite{agr}); the EMOS and EPN
spectra were fitted simultaneously with the temperature and abundance
linked. The EPN normalisation was not linked to that of the EMOS
cameras (it is typically slightly lower).  

We show the resulting projected abundance profiles in
Fig.~\ref{fig:abund} (the temperature profiles are discussed further
below). There is strong line emission from several elements at the
average temperature of these clusters, which will be investigated in
further detail in a future paper.  

\subsubsection{Correction for Projection and PSF effects}
\label{sec:deprojpsf}

\begin{figure}
\begin{centering}
\includegraphics[scale=1.,angle=0,keepaspectratio,width=\columnwidth]{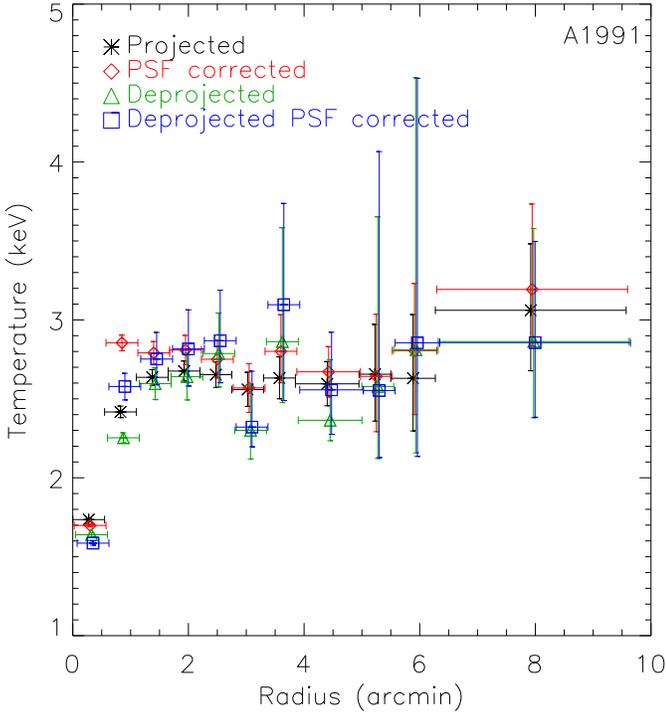}
\caption{{\footnotesize  An illustration of the effects of PSF
blurring and projection. The projected temperature profile of A1991
(stars) is shown here compared to the PSF-corrected profile
(diamonds), the deprojected profile (triangles) and the PSF-corrected,
deprojected profile (squares). The effects play a subtle role in the
central regions, but the exterior temperature extimates are fully
compatible with the projected profile. Errors are $1\sigma$ confidence
level. {\it The electronic edition of the journal contains a colour
version of this Figure.}}}\label{fig:a1991tprof}  
\end{centering}
\end{figure}

The projected temperature profiles show evidence for central cool
regions of varying strength in all the clusters, the most striking
example of which is A1991. It is clear that both PSF and projection
effects may play a role, and will need to be taken into account if we
want to recover the best estimate of the mass profile. For the
following analysis we use the method of Pointecouteau \etal
(\cite{pointeco}), where the observed annular spectra, $S^O_i (E)$,
are modelled with a linear combination of absorbed isothermal {\sc
MEKAL} models: 

\begin{equation}
S^O_i (E) = {\rm WABS} (N_H^i)   \sum_{j=1}^n a_{i,j} {\rm MEKAL}
(T_j, Z_j).\label{eqn:redist} 
\end{equation}

For pure PSF correction, the $a_{i,j}$ redistribution coefficient is
the emission measure contribution of ring $j$ to ring $i$, where the
coefficients were calculated using the emission measure profile from
the best-fitting gas density profile model, convolved with the \xmm\
PSF at 1 keV. The redistribution coefficients for pure projection
effects are calculated as the emission measure contribution of the
shell $j$ to the ring $i$. PSF and projection effects are taken into
account using the emission measure contribution of the shell $j$ to
the ring $i$ after convolution with the PSF. 

For each cluster the annular spectra were fitted simultaneously in
{\sc XSPEC}. We fixed the absorption to the galactic value, and we
froze the abundance of each {\sc MEKAL} model to the best-fit value
found for each projected annulus. For $n$ annuli, the free parameters
are thus the $n$ temperatures and $n$ normalisations, one per annulus,
the other normalisations being linked according to
Eq.~\ref{eqn:redist}. Further details of the fitting procedure can be
found in Pratt \& Arnaud~(\cite{pa02}) and Pointecouteau
\etal~(\cite{pointeco}). 

Figure~\ref{fig:a1991tprof} illustrates the separate effects of PSF
blurring and projection on the the temperature profile of A1991
(results for A2717 and MKW9 are similar).  
There is a definite effect in the central regions.  While in both
cases the temperature of the central bin is decreased, the PSF
correction and deprojection have an opposite effect on the second
bin. Beyond the third annulus ($\sim 2\arcmin$) the corrected profiles
are consistent with the projected profile. This is expected, the
external profile being very flat. However, we can see that the
correction process amplifies considerably the slightest variation in
the profile and the corrected profile is noisier.  

The noise increase is particularly marked for the deprojected, PSF
corrected profile, where the largest deviations correspond to
projected annuli which have slightly lower or higher temperatures than
the best fit  smooth profile. Similar behaviour was also seen in
A2717, MKW9 and in the analysis of A478 by Pointecouteau et
al.~(\cite{pointeco}). These deviations are very likely an artifact of
the correction process, and as such should not be considered as
physically meaningful. In the central regions, the deprojected, PSF
corrected temperatures are subtly different from the projected
values. 

\begin{figure}
\begin{centering}
\includegraphics[scale=1.,angle=0,keepaspectratio,width=\columnwidth]{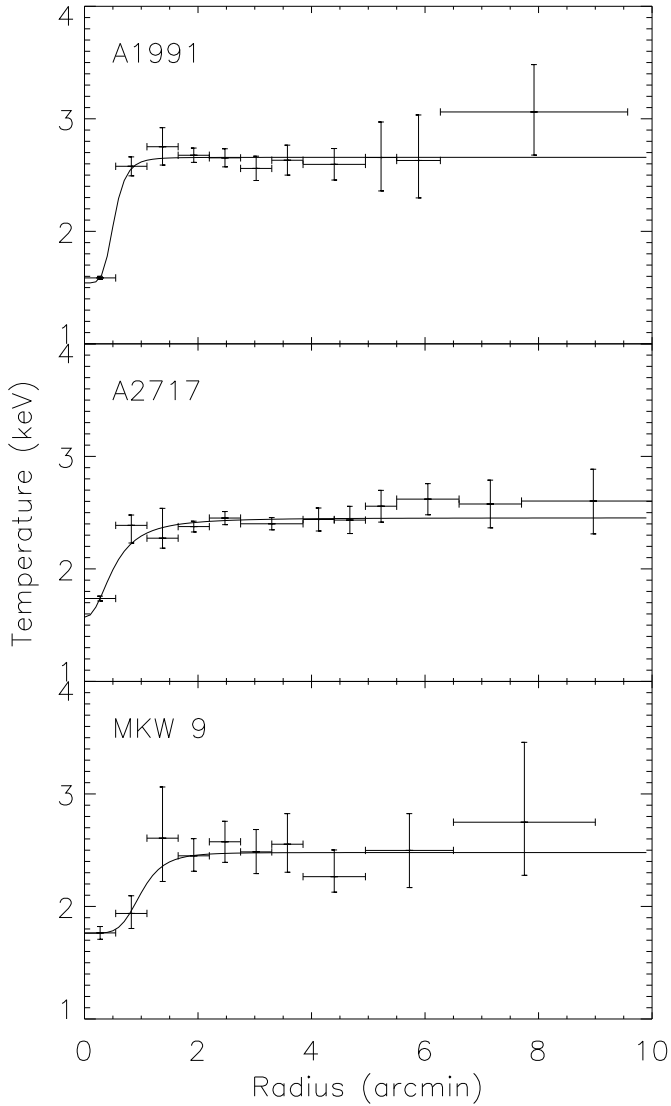}
\caption{{\footnotesize Cluster temperature profiles. These composite
profiles are combined from the inner three annuli of the deprojected,
PSF corrected profile, plus the projected temperature values
thereafter (see discussion in Sect.~\ref{sec:deprojpsf}). Errors are
$1\sigma$. The solid lines are the best fits to each profile with a
function of the form given in Eq.~\ref{eqn:allen}; fit parameter
values are given in Table~\ref{tab:ktfit}.}}\label{fig:tprofs} 
\end{centering}
\end{figure}

Since for all the clusters the outer regions of the deprojected, PSF
corrected temperature profiles are i) subject to unphysical jumps
which would lead to mass discontinuities, but ii) consistent with the
projected profile, for the following analysis we use a composite
temperature profile made up of the inner three annuli of the
deprojected, PSF corrected profile, plus the projected temperature
values thereafter. These composite profiles are shown in
Fig.~\ref{fig:tprofs}. 

\subsubsection{Temperature profile modelling}
 We also modelled the composite temperature profile of each cluster
discussed above (corrected for projection and PSF effects) with the
function described in Allen et al.~(\cite{asf01}), viz., 
\begin{equation}
T = T_{\rm 0} +T_{\rm 1}  [(r/\rc)^{\eta} / (1 +
(r/\rc)^{\eta})],\label{eqn:allen} 
\end{equation}
The parameters for each cluster are given in Table~\ref{tab:ktfit}.

%
\begin{table}
\begin{minipage}{\columnwidth}
\caption{{\footnotesize Results of the best fit of the deprojected,
PSF-corrected temperature profiles (see Sect.~\ref{sec:deprojpsf})
with the analytical model given by Eq.~\ref{eqn:allen}.}} 
\label{tab:ktfit}
\centering
\begin{tabular}{l l l l l}
\hline
\hline
Cluster & $T_{\rm 0}$& $T_{\rm 1} $ & $\rc$ & $\eta$\\
 & (keV) & (keV)\\ 
\hline
A1991 & $1.54$ & $1.12$ &0\farcm 52& 5.\\ 
A2717 &  $1.57$ & $0.88$ &0\farcm 52& 2.28\\ 
MKW9  &  $1.76$ & $0.72$ &1\farcm 00& 5.\\ 
\hline
\end{tabular}
\end{minipage}
\end{table}

\subsection{Virial temperature}
\label{sec:globspec}

\begin{table}
\begin{minipage}{\columnwidth}
\caption{{\footnotesize The temperature and abundance values found
after simultaneous fits of the EMOS and EPN global spectra of each
cluster with an absorbed {\sc MEKAL} model. The global spectrum was
extracted in the radial range $0.1~\rv \leq r \leq 0.3~\rv$, where
$\rv$ comes from the best-fitting NFW model discussed later in the
text. The abundances are relative  to Anders \&
Grevesse~(\cite{agr}). Errors are $1\sigma$ for one interesting
parameter.}}\label{tab:glob} 
\centering
\begin{tabular}{l l l }
\hline
\hline
Cluster & $\kT$& $Z$\\
 & (keV) & ($Z_\odot$)\\ 
\hline
A1991 &  
        $2.65^{+0.05}_{-0.05}$ & $0.33^{+0.03}_{-0.02}$ \\ 
A2717 & 
        $2.53^{+0.05}_{-0.05}$ & $0.34^{+0.02}_{-0.02}$ \\ 
MKW9  & 
        $2.58^{+0.15}_{-0.15}$ & $0.37^{+0.07}_{-0.06}$ \\ 
\hline
\end{tabular}
\end{minipage}
\end{table}

To compare the properties of these cool clusters with those of hotter
systems (e.g. Sect.~\ref{sec:entropy}) it is useful to define a global
temperature, representative of the `virial' temperature.  Faced with
the considerations i) that the measurements must be easily
reproducible both for observers and simulators alike, ii) that in none
of these observations are we able to detect emission much beyond $\sim
0.4~\rv$ and iii) that all clusters appear to have somewhat cooler gas
in the core region, we extracted a spectrum for each cluster from all
events between $0.1~\rv \leq r \leq 0.3~\rv$. The $\rv$ in each case
comes from the best-fitting NFW mass model to the mass profiles,
discussed in more detail in Sect.~\ref{sec:massmodel}. 

The EMOS and EPN spectra of each cluster were fitted  simultaneously
in the [0.3 - 6.0] keV band with an absorbed {\sc MEKAL} model.  Fits
were identical to those described in Sect.~\ref{sec:globspec}. 
The results of this global analysis are given in Table~\ref{tab:glob}.  

The temperature of A1991 is in excellent agreement with the
temperature $\kT \sim 2.7$ keV, derived with Chandra beyond the
cooling core (Sharma \etal ~\cite{sharma}). The XMM temperature of
A2717 is higher than the temperature $\kT = 1.6 \pm 0.3 $ keV derived
by Liang et al. (\cite{liang}). This value was found from an
isothermal fit to the ROSAT data within $8\arcmin$, actually an
emission weighted temperature including the cooling core emission, and
is thus likely to be biased low (especially since ROSAT is
particularly sensitive to any cool component). The global temperature
for MKW9 is in good agreement with previous ASCA studies by e.g.,
Finoguenov et al.~(\cite{fad01}; $\kT = 2.92 \pm 0.43$ keV) and
Sanderson et al.~(\cite{sand03}; $\kT = 2.88^{+0.68}_{-0.55}$ keV). It
is also in agreement with the analysis of the same \xmm\ observation
by Kaastra et al. (\cite{kaa}).  

\section{Mass profiles}
\label{sec:mass}

\subsection{Mass profile calculation}

Combining the gas density (Sect.~\ref{sec:gasden}) and temperature
(Sect.~\ref{sec:deprojpsf}) profiles, we can derive a total
gravitational mass profile under the assumptions of hydrostatic
equilibrium and spherical symmetry. The mass was calculated at each
radius of the temperature profile using the adapted version of the
Monte Carlo method of Neumann \& B\"ohringer~(\cite{nb95}) described
in Pratt \& Arnaud~(\cite{pa03}). This takes as input the parametric
model for the gas density profile and the measured temperature profile
with errors. A random temperature is calculated at each radius of the
measured temperature profile, assuming a Gaussian distribution with
sigma equal to the $1\sigma$ error, and a cublic spline interpolation
is used to compute the derivative. Only profiles corresponding to a
monotonically increasing mass gradient are kept: 1000 such profiles
were calculated. The input temperature profile is the composite
profile discussed in Sect.~\ref{sec:deprojpsf}. 

The uncertainty in the modelling of the gas density profile was taken
into account by calculation of the errors on the density gradient at
each point. The final uncertainties on each mass profile point are
then the quadratic addition of these errors with the Monte Carlo
errors.  

\begin{figure}
\begin{centering}
\includegraphics[scale=1.,angle=0,keepaspectratio,width=\columnwidth]{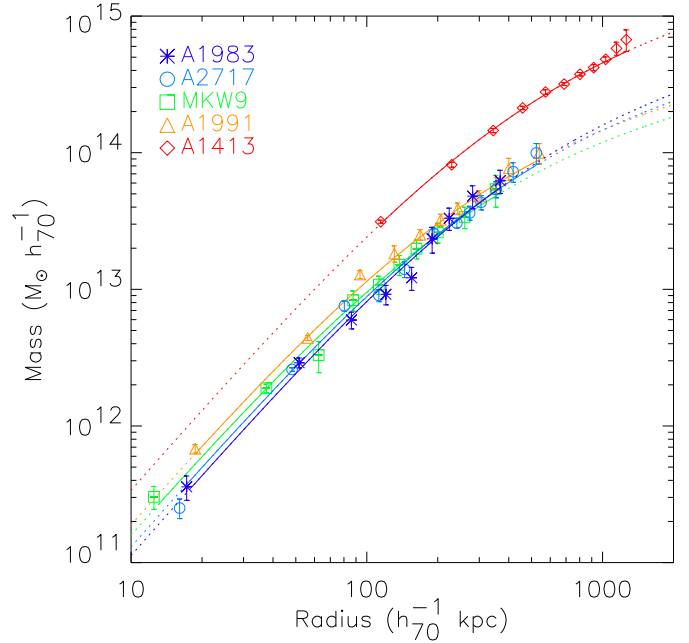}
\caption{{\footnotesize Integrated total gravitating mass profiles,
shown with $1\sigma$ errors. The solid lines are the best-fitting NFW
profiles as detailed in Table~\ref{tab:mprof}. Dotted lines represent
extrapolations of the best-fitting NFW models. The data for  A1983 and
A1413 are from Pratt \& Arnaud (\cite{pa03}) and Pratt \& Arnaud
(\cite{pa02}) scaled to the $\Lambda$CDMH70
cosmology. {\it See the electronic edition of the Journal for a colour
version of this figure.}}}\label{fig:unscmprofs}   
\end{centering}
\end{figure}

\subsection{Mass profile modelling}
\label{sec:massmodel}

The integrated mass profile of each cluster was then fitted using the
Navarro et al.~(\cite{nfw97}; hereafter NFW) density distribution
$\rho(r) \propto [(r/\rs) (1+r/\rs)]^{-1}$. This model has two free
parameters: a normalisation factor and the scaling radius $\rs$, or
equivalently the total mass $M_{200}$ and the concentration parameter
$c_{200} = r_{200}/\rs$ (see e.g., Suto, Saski \& Makino  \cite{suto}
for details). $M_{200}$ is defined as the mass contained in a sphere
of radius $r_{200}$, which encompasses a mean density  of 200 times
the critical density at the cluster redshift $\rho_{\rm c}(z)= 3
h(z)^2{\rm H_0} /8\pi{\rm G}$, where $h^{2}(z)=\Omega_{\rm m}(1+z)^{3}
+\Omega_{\Lambda}$ . In numerical simulations, this sphere is found to
correspond roughly to the virialised part of clusters. Results of
these fits are shown in
Table~\ref{tab:mprof}. Fig.~\ref{fig:unscmprofs} shows the cluster
mass profiles plotted in physical units, overlaid with the
best-fitting NFW models.   

We also tried fitting these mass profiles with an isothermal sphere
model and a Moore \etal (\cite{mqgsl99}) model: the former was not an
acceptable fit to any of these clusters, while the latter was
unconstrained. 

\begin{table}
\caption{{\footnotesize Results from the NFW fits to the mass profiles.}}\label{tab:mprof}
\centering
\begin{tabular}{l l l l }
\hline
\hline
Parameter  & A1991&A2717 & MKW9\\
\hline
$\Lambda$CDMH70 \\
\cline{1-1}
$c_{200}$                    & $5.7^{+0.4}_{-0.3}$        &
$4.2^{+0.3}_{-0.3}$ & $5.4^{+0.7}_{-0.7}$ \\ 
$\rs$ ( kpc)      & $191^{+19}_{-17}$ & $261^{+27}_{-24}$ &
$186^{+45}_{-34}$ \\ 
$\rv$ ( kpc)  & 1105 & 1096 & 1006 \\ 
$M_{200}$ ($10^{14} M_{\odot}$) & 1.63 & 1.57 & 1.20 \\ 
\\ 
$\chi^2 / \nu$                  & 9.98/9         & 15.8/10    & 4.0/8
\\ 
\hline
SCDMH50\\
\cline{1-1}
$c$                            & $5.6^{+0.4}_{-0.3}$ &
$4.1^{+0.3}_{-0.2}$ & $5.3^{+0.7}_{-0.7}$ \\ 
$\rs$ (kpc)      & $260^{+26}_{-23}$  & $358^{+37}_{-33}$  &
$255^{+61}_{-46}$   \\ 
$\rv$ (kpc)  & 1466          & 1466       &  1358  \\
$M_{200}$ ($10^{14} M_{\odot}$) & 2.17           & 2.12        & 1.63  \\
$\chi^2 / \nu$                  & 9.98/9         & 15.8/10    & 4.0/8  \\
\hline
\end{tabular}
\end{table}

\subsection{Cluster dynamical state and NFW fit}
The NFW model is not a very good fit to the A2717 mass profile (the
$\chi^2_{\nu} \sim 1.6$).  Jing~(\cite{jing00}) has shown that
non-equilibrium haloes produce poor fits to the NFW distribution, and
also that the concentration parameters are generally lower in these
cases. A2717 has the lowest $c$ and the worst-fitting NFW profile, and
so this may well be due to the halo being unrelaxed.  An optical
substructure study by Girardi et al.~(\cite{getal97}) has suggested
that there is a significant structure in the central region, slightly
foregrounded by $\sim 600$ km s$^{-1}$, and so the core region may
indeed be dynamically perturbed.  However, we do not see any obvious
structure either in the X-ray/optical overlay or in the hardness ratio
map.  Moreover, A1991 and MKW9 are well fitted with the NFW profile
and have similar $c$ values, but while A1991 appears  relaxed, MKW9
does not. The isophotal twisting from the centre to the outer regions
(Fig.~\ref{fig:optxim}), and the considerable structure in the
hardness ratio image (Fig.~\ref{fig:hrim}), together with the evidence
for substructure from optical studies (Beers et al.~\cite{beers}), all
suggest that MKW9 may not be fully relaxed. 

\begin{table}
\caption{{\footnotesize A comparison of the dark matter velocity
dispersions calculated from the $\sigma_{DM}$--$M_{200}$ relation of
Evrard \& Gioia~(\cite{eg}; Eq.~\ref{eqn:eg}), with the
optically-derived galaxy velocity dispersions. Values are given in km
s$^{-1}$.}}\label{tab:sigprof} 
\begin{minipage}{\columnwidth}
\centering
\begin{tabular}{l l l l }
\hline
\hline
\multicolumn{1}{l}{Cluster} & \multicolumn{1}{c}{$\sigma_{DM}$} & \multicolumn{1}{l}{$\sigma_{\rm opt}$} & \multicolumn{1}{l}{Reference}  \\
\hline
A1991 & $526$ &  $631^{+147}_{-137}$ & Girardi et al. (\cite{getal97})\\
A2717 & $520$ &  $541^{+65}_{-41}$   & Girardi et al. (\cite{getal97})\\
MKW9  & $474$ &  $579^{+331}_{-337}$ &  Beers et al. (\cite{beers})\\
\hline
\end{tabular}
\end{minipage}
\end{table}

We can use the tight correlation between $M_{200}$ and the velocity
dispersion of the dark matter found in the Hubble Volume simulations
(Evrard \& Gioia~\cite{eg}), viz.,  
\begin{equation}
\sigma_{DM} = 1075
[h(z)~M_{200}/(10^{15}~h_{100}^{-1}~M_{\odot})]^{1/3}~{\rm km~s}^{-1},
\label{eqn:eg} 
\end{equation}
\noindent 
to further probe the link, if any, between the NFW fits to these
clusters and their dynamical state. The optically-derived galaxy
velocity dispersion is a measure of the actual  dark matter  velocity
dispersion, both collisionless components presumably following the
same dynamics.  In Table~\ref{tab:sigprof}, we compare  the
$\sigma_{DM}$ values expected from the best-fitting NFW mass models to
these clusters with the galaxy velocity dispersions.  For unrelaxed
clusters we would not expect good agreement because of i) incorrect
estimates of the total cluster mass through the HE equation and/or ii)
intrinsic departure from the $M_{200}$--$\sigma$ relation.  We can see
that not only are the results in excellent agreement\footnote{ In view
of the large error bars on the optical velocity, the agreement
unfortunately does not give stringent constraints on the dynamical
sate of  MKW9. }, but that the agreement is particularly good for
A2717. This gives us confidence in the total mass estimates derived
from the NFW fits to these X-ray derived mass profiles, and suggests
that none of these clusters can be too far from equilibrium. 

That being said, it is highly ulikely that the three clusters are in
exactly the same state of relaxation. Judging by the X-ray and optical
information, MKW9 is probably the least relaxed. In this sample,
though, there is no obvious link between the dynamical state and the
goodness of the NFW fit or the value derived for the concentration
parameter.  

\begin{figure}[t]
\begin{centering}
\includegraphics[scale=1.,angle=0,keepaspectratio,width=\columnwidth]{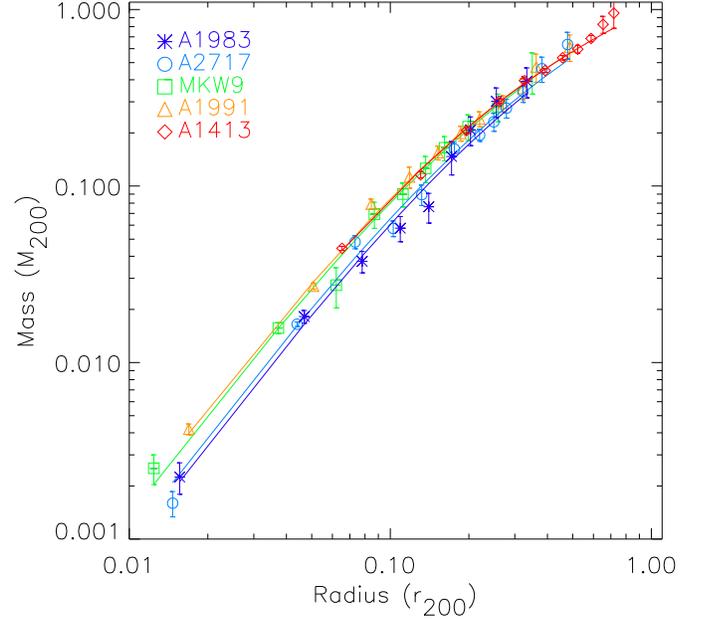}
\caption{{\footnotesize Cluster mass profiles scaled in units of
$r_{200}$ and $M_{200}$. $r_{200}$ and $M_{200}$ were calculated from
the best fitting NFW model (Sect.~\ref{sec:massmodel}). {\it See the
electronic edition of the Journal for a colour version of this
figure.}}}\label{fig:mprofs}  
\end{centering}
\end{figure}

\subsection{Scaled mass profiles}

We now turn our attention to the scaled mass profiles of these
systems, where we scale the radius by $r_{200}$ and the mass by
$M_{200}$, these values coming from the best fitting NFW model to each
cluster.  Comparison of these scaled profiles allows us to assess
similarity in the shape of the mass profiles. To increase the sample
and allow a first comparison with massive clusters, we also add data
from our previously published work on the slightly cooler cluster
\object{A1983} (Pratt \& Arnaud~\cite{pa03}) and the hot, massive
cluster \object{A1413} (Pratt \& Arnaud~\cite{pa02}). The mean
temperature in the radial range $0.1 \rv - 0.3 \rv$ is $\kT = 2.2$ keV
and $\kT = 6.5$ keV for A1983 and A1413, respectively. The mass
profiles of these two clusters\footnote{The published data  are scaled
to the $\Lambda$CDMH70 cosmology.} are plotted in physical units in
Fig.~\ref{fig:unscmprofs}.  The scaled mass profiles of all five
clusters ($\Lambda$CDMH70) are shown overlaid with the best fit scaled
NFW models in Figure~\ref{fig:mprofs}.   

\begin{table}
\begin{minipage}{\columnwidth}
\caption{{\footnotesize Relative dispersion in scaled mass  profiles,
measured using the standard deviation and mean (first and third
columns) or the biweight estimators of Beers et
al.~(\cite{beers1}). }}\label{tab:dispm} 
\centering
\begin{tabular}{l cccc}
\hline
\hline
\multicolumn{1}{l}{Radius}  & \multicolumn{2}{c}{$\Lambda$CDMH70} & \multicolumn{2}{c}{SCDMH50} \\ 
\multicolumn{1}{l}{ } & $\sigma/m$ & $S_{\rm BI}/C_{\rm BI}$ & $\sigma/m$ & $S_{\rm BI}/C_{\rm BI}$\\ 
\hline
\multicolumn{3}{l}{Scaled mass: NFW best fit  model} \\
\cline{1-3}
$0.05~\rv$  & $0.18$ & $0.19$ & $0.18$ & $0.19$\\
$0.1~\rv$  & $0.15$ & $0.16$ & $0.15$ & $0.16$\\
$0.3~\rv$ & $0.08$ & $0.09$ & $0.08$ & $0.08$\\
$0.5~\rv$ & $0.04$ & $0.05$ & $0.04$ & $0.04$ \\
\hline
\multicolumn{3}{l}{Scaled mass:  interpolated data} \\
\cline{1-3}
$0.05~\rv$  & $0.18$ & $0.20$ & $0.17$ & $0.18$\\
$0.1~\rv$ & $0.25$ & $0.25$ & $0.24$ & $0.24$\\
$0.3~\rv$ & $0.06$ & $0.02$ & $0.06$ & $0.04$ \\
$0.5~\rv$ & $0.13$ & $0.13$ & $0.13$ & $0.12$\\
\hline
\end{tabular}
\end{minipage}
\end{table}

It is clear from Fig.~\ref{fig:unscmprofs} that the mass profiles of  A1991, A2717 and MKW9 are genuinely similar even in physical units. When we add the data from  A1983 and A1413 and scale the profiles, the resemblance is remarkable. To further quantify this similarity,  we have estimated the relative dispersion in the scaled mass profiles at 0.05, 0.1, 0.3 and $0.5~\rv$.  The scaled masses at these radii were estimated using either i) the best-fitting NFW mass model or ii) a linear inter/extrapolation of the measured $M(r)$ in the $\log$--$\log$ plane. In the case of A1983 and MKW9, there is a small amount of extrapolation needed to reach $0.5~\rv$.  We used both the standard deviation and mean and the biweight estimators for location and scale ($C_{BI}$ and $S_{BI}$), described in Beers et al.~(\cite{beers1}), to estimate the relative dispersion.
The resulting values are shown in Table~\ref{tab:dispm}. The scaled
mass profiles depend on the cosmological model through the function
$\rho_{\rm c}(z)$, used in the definition of $\rv$ and $M_{200}$, and
via the angular distance used to convert the angular radius to
physical units. The cool systems are all at roughly the same redshift
but A1413 is at a somewhat higher redshift of $z=0.14$. We thus also
give the dispersions obtained for the SCDMH50 cosmology in
Table~\ref{tab:dispm}.  

\begin{figure}
\begin{centering}
\includegraphics[scale=1.,angle=0,keepaspectratio,width=\columnwidth]{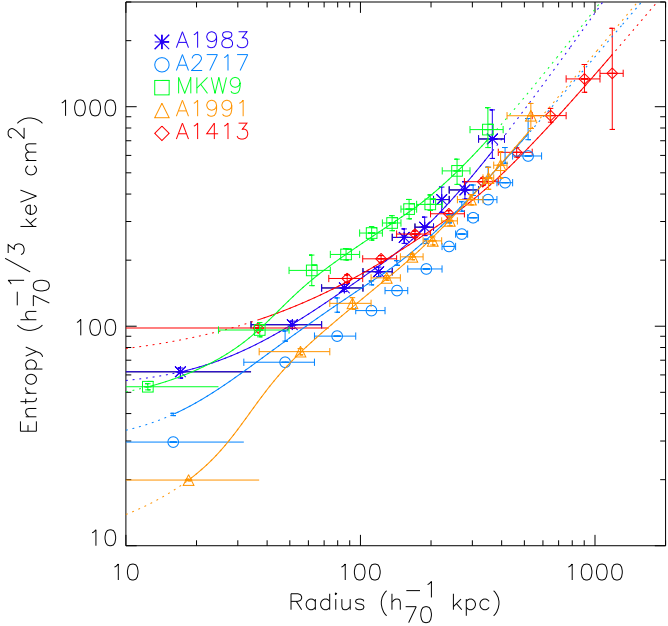}
\caption{{\footnotesize Cluster entropy ($S = \kT \ne^{-2/3}$)
profiles, derived from the deprojected and PSF corrected temperature
profiles and the best-fitting analytical model for the gas
density. The solid lines represent the entropy profiles obtained using
analytic models for both the gas density and temperature
distributions. Dotted lines represent extrapolations of these models.
The data for  A1983 and A1413 are from Pratt \& Arnaud
(\cite{pa03}). {\it See the electronic edition of the Journal for a
colour version of this figure.}}}\label{fig:unsceprofs}   
\end{centering}
\end{figure}

As can been seen from Table~\ref{tab:dispm}, for the present sample,
the dispersions are neither sensitive to the cosmology nor to the
choice of the statistical estimator.  The best fitting scaled NFW
profile of each cluster depends only on the concentration
parameter. By definition of the scaling the dispersion is null at
$r=\rv$, and naturally increases with decreasing scaled radius. The
observed dispersion is small, less than  $\sim 15$ per cent  for  $r >
0.1~\rv$, reflecting  the modest dispersion in the concentration
parameter values (further discussed in Sect.~\ref{sec:disdm}). Turning
now to the interpolated data, the dispersion is generally higher
(e.g. $25$ per cent at $0.1 \rv$) and its variation with radius is
more chaotic. This is expected, since in that case statistical errors
on the data contribute to the dispersion. The dispersion obtained
using the best fitting NFW models is probably more representative of
the intrinsic dispersion in the mass profiles.  A more rigorous
estimate of this intrinsic dispersion would require a much larger
cluster sample and is beyond the scope of this paper.  

\section{Entropy}
\label{sec:entropy}

\subsection{Entropy profiles}
We next determined the gas entropy ($S = \kT \ne^{-2/3}$) profiles for
the clusters, using the analytic description for the gas density
profile and the observed temperature profile corrected for PSF and
projection effects (Sect.~\ref{sec:deprojpsf}).  The resulting
profiles, together with the  profiles of A1983 and A1413 (Pratt \&
Arnaud~\cite{pa03}), are shown plotted in physical units in
Fig.~\ref{fig:unsceprofs}. The profiles are shown with typical errors
corresponding to the error on each temperature bin.   
We also calculated the entropy using the analytical temperature
profile model (Eq.~\ref{eqn:allen}).  These profiles are also plotted
in the Figure.  

\begin{figure}
\begin{centering}
\includegraphics[scale=1.,angle=0,keepaspectratio,width=\columnwidth]{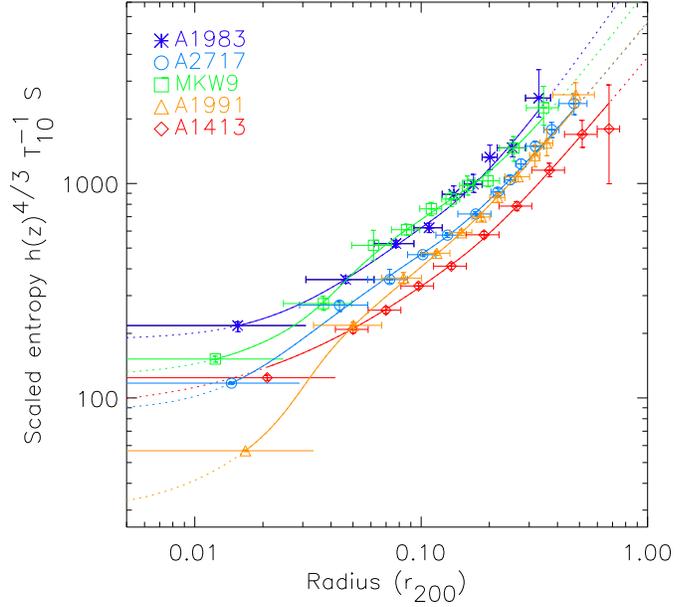}
\caption{{\footnotesize Scaled entropy profiles. The radius is scaled
to $\rv$ measured from the best fitting NFW mass model. The entropy is
scaled following the prediction of the standard self-similar model of
cluster formation: $S \propto h(z)^{-4/3} T$ . The scaling is
performed using the global temperature  $\Tx$, estimated as described
in Sect.~\ref{sec:globspec} in units of $10 \keV$. {\it See the
electronic edition of the Journal for a colour version of this
figure.}}}\label{fig:e1} 
\end{centering}
\end{figure}
\begin{figure}
\begin{centering}
\includegraphics[scale=1.,angle=0,keepaspectratio,width=\columnwidth]{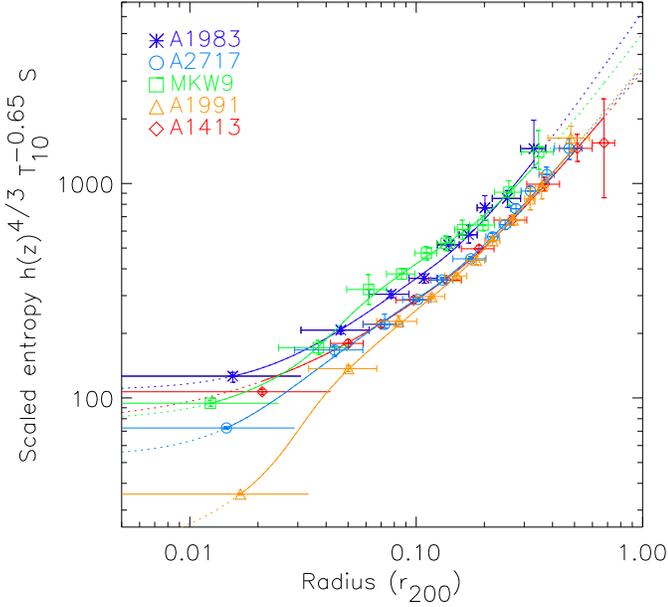}

\caption{{\footnotesize Same as Fig.\ref{fig:e1} but the empirical
scaling $S \propto h(z)^{-4/3} T^{0.65}$  is used instead. This
scaling significantly decreases the dispersion between the
profiles. {\it See the electronic edition of the Journal for a colour
version of this figure.}}}\label{fig:e065}   
\end{centering}
\end{figure}

\begin{table}
\begin{minipage}{\columnwidth}
\caption{{\footnotesize Relative dispersion in scaled  entropy
profiles, measured using the standard deviation and mean (first and
third columns), or the biweight estimators of Beers et
al.~(\cite{beers1}). }}\label{tab:dispe} 
\centering
\begin{tabular}{l cccc}
\hline
\hline
\multicolumn{1}{l}{Radius}  & \multicolumn{2}{c}{$\Lambda$CDMH70} &
\multicolumn{2}{c}{SCDMH50} \\  

\multicolumn{1}{l}{ } & $\sigma/m$ & $S_{\rm BI}/C_{\rm BI}$ &
$\sigma/m$ & $S_{\rm BI}/C_{\rm BI}$\\  
\hline
\multicolumn{3}{l}{Scaled Entropy:  $T^{-1}$ scaling} \\
\cline{1-3}
$0.05~\rv$ & 0.30 &0.28 & 0.28& 0.26\\
$0.1~\rv$ & 0.30 &0.29 & 0.29& 0.28\\
$0.3~\rv$ & 0.30 &0.29 & 0.28& 0.26 \\
$0.5~\rv$ & 0.34& 0.33 & 0.31 & 0.31 \\
\hline
\multicolumn{3}{l}{Scaled Entropy:  $T^{-0.65}$ scaling} \\
\cline{1-3}
 $0.05~\rv$ & 0.22 &0.22 & 0.21& 0.20 \\
 $0.1~\rv$ & 0.22 &0.24 & 0.21& 0.21 \\
$0.3~\rv$ & 0.20 &0.26 & 0.18& 0.20 \\
$0.5~\rv$ & 0.24 &0.29 & 0.21 &0.19 \\
\hline
\end{tabular}
\end{minipage}
\end{table}

\subsection{Scaled entropy profiles}

In the standard self-similar scenario, clusters form at constant
density contrast and the gas simply follows the dark matter. The mean
gas and DM density are proportional to the critical density of the
Universe: $\overline{\ne} \propto \overline{\rhoDM} \propto \rhocz
\propto h^{2}(z)$. The entropy thus scales as $S \propto h(z)^{-4/3}
T$, and the scaled entropy profiles ($h(z)^{4/3} T^{-1} S$ versus
$r/\rv$) of all clusters should coincide.   We show the entropy
profiles with this scaling in  Fig.~\ref{fig:e1}, where $\Tx$ is the
global temperature estimated as described in Sect.~\ref{sec:globspec}
in units of $10 \keV$ (hereafter $T_{10}$). Recent results (Ponman et
al.~\cite{psf03}) suggest that entropy scales with temperature such
that $S \propto \Tx^{0.65}$.  
The entropy profiles with this scaling are shown in
Fig.~\ref{fig:e065}. 

As with the mass profiles, we have inter/extrapolated the entropy
profiles to 0.05, 0.1, 0.3 and $0.5~\rv$ using the best-fitting
(analytical) gas density and temperature models for each cluster. The
dispersion in the scaled profiles was once again calculated for the
two cosmologies using both the standard and  biweight estimators. The
results are shown in Table~\ref{tab:dispe}.  If the standard scaling
is used the dispersion is about $\sim 30$ per cent, and is insensitive
to the choice of the cosmology or statistical estimator. There is less
scatter in the scaled entropy profiles if one uses the
empirically-determined scaling of Ponman et al. (\cite{psf03}), $S
\propto T^{0.65}$. This is obvious if one compares Fig.~\ref{fig:e1}
and Fig.~\ref{fig:e065}. The improvement is quantified in
Table~\ref{tab:dispe}: the typical dispersion is $\sim 22$ per cent, a
drop of $\sim 40- 50$ per cent from that of the standard scaling (the
drop is somewhat smaller for the biweight estimator).

The dispersion is remarkably constant with radius (a variation of less
than $10$ per cent),  indicating strong similarity in the shape of the
entropy profile. This is also clear from Fig.~\ref{fig:e065}:  while
there is some variation of form in the inner regions,  beyond about
$0.05-0.1~\rv$, the profiles all become approximately parallel. This
suggests that the form of $\kT \gtrsim 2$ keV cluster entropy profiles
is not temperature dependent.  The shape, which is essentially
governed by the shape of the density profile, is not exactly a
powerlaw: it flattens slightly with decreasing radius following the
double $\beta$--model.  However, within the radial range
$[0.05-0.5]\rv$ the mean scaled profile is well approximated by a
powerlaw: 

\begin{equation}
h(z)^{4/3} T_{10} ^{-0.65} S (r) =  470 \left(\frac{r}{0.1\rv}\right)^{0.94\pm0.14} \hs^{-1/3}~\keV~{\rm cm^{2}},
\label{eq:sr}
\end{equation}

where the error on the slope is derived taking into account the
$\pm1\sigma$ dispersion around the mean profile.  The slope is close
to, but slightly shallower than, the $S \propto r^{1.1}$ behaviour
expected from analytical modelling of shock heating in spherical
collapse (Tozzi \& Norman~\cite{tn01}), behaviour which is also seen
in cosmological simulations (e.g., Borgani et al.~\cite{borg02}). This
is further discussed in Sec.~\ref{sec:disgas}.  

The fact that Figures~\ref{fig:unsceprofs} (unscaled profiles)
and~\ref{fig:e065} (profiles scaled with the $S \propto T^{0.65}$
relation and the virial radius) show similar scatter is the result of
a close coincidence in logarithmic slopes. For $r_{200} \propto
T^{0.57}$ (corresponding to $M \propto T^{1.7}$; e.g, Finoguenov et
al~\cite{frb}), and $S \propto T^{0.65}$, the $\log{S}$--$\log{r}$
curves are translated along a line of slope $-0.65/-0.57 = 1.14$,
which is similar to observed radial variation of $S(r) \propto 0.94$.
The scaled profile of A1413 (the hottest cluster) is translated a
small bit lower than those of the cool clusters (since $0.65 > 0.57$).
The effect of scaling with redshift then compensates a small amount
for this.

\section{Discussion}
\label{sec:dis}

\subsection{The gravitational collapse of the dark matter}
\label{sec:disdm}

The observed shape of the mass profile is an important test of our
understanding of the dark matter collapse.   Although the exact slope
of the profile in the very centre of clusters is still a matter of
debate, all numerical simulations of structure formation predict a
universal form with a central cusp (e.g. Navarro \etal~\cite{nfw04},
and references therein).  

The NFW profile provides the best fit to the three low mass clusters
studied here, and a King profile (i.e. a profile with a core) is
rejected.   These three poor clusters thus display the cusped
distribution expected from numerical simulations of CDM collapse, in
common with A1983 (Pratt \& Arnaud~\cite{pa03}) and with hotter
clusters observed with  \xmm or \chandra  (e.g., David et
al.~\cite{david01}; Allen et al.~\cite{asf01}; Arabadjis et
al.~\cite{abg02}, Pratt \& Arnaud~\cite{pa02}; Lewis et
al.~\cite{lewis}; Pointecouteau et al.~\cite{pointeco}; Buote \&
Lewis~\cite{bl04}).  The profiles we have derived are not precise
enough for us to constain the exact slope in the centre, although we
note that the few observations (of massive clusters ) precise enough
to constrain this slope favour an NFW type-profile (Lewis et
al.~\cite{lewis}; Pointecouteau et al.~\cite{pointeco}; Buote \&
Lewis~\cite{bl04}).  

There is a remarkable similarity between the mass profiles of the cool
systems and that of the hot cluster A1413.  The dispersion in the
scaled profiles is less than $20$ per cent in the radial range  $0.05
\rv - 0.5 \rv$, where the various profiles could be compared without
excessive extrapolation.  In other words, the shape of the mass
profiles from low mass systems up to high mass systems is very close
to Universal.  We now address the question of whether this shape,
which is defined by the concentration parameter,  is {\it
quantitatively} consistent with the predictions.  

\begin{figure}
\begin{centering}
\includegraphics[scale=1.,angle=0,keepaspectratio,width=\columnwidth]{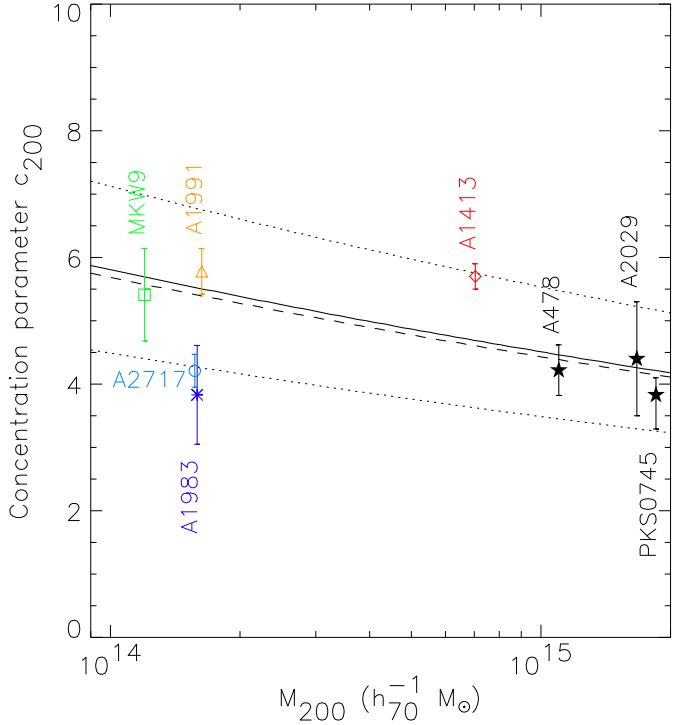}
\caption{{\footnotesize Variation of the concentration parameter
$c_{200}$ with cluster mass $\Mv$. The data points are a compilation
of the present work (A2717, A1991 and MKW9) and data from the
literature. We have only considered local clusters ($z<0.15$) with
$c_{200}$ measurements precise to better than $\pm 20$ per cent. The
data from  A1983,  A1413, A478, A2029 and PKS0745-191 are from Pratt
\& Arnaud (\cite{pa03}), Pratt \& Arnaud (\cite{pa02}), Pointecouteau
et al. (\cite{pointeco}),  Lewis et al. (\cite{lewis}) and  Allen et
al. (\cite{asf01}), respectively. The published PKS0745-191 values
correspond to a density contrast of 2500, these were converted to
$c_{200}$ and $\Mv$ using the published best fit NFW model. All errors
are $1\sigma$  errors.  Lines: theoretical variation of the mean
concentration with mass at $z=0$ (full line) and $z=0.15$ (dashed
line) from the numerical simulations of Dolag \etal
(\cite{dol04}). The published relation c(M) was translated to the
virial radius convention used here (see text for details). The dotted
lines correspond to the mean plus/minus the standard
deviation. {\it See the electronic edition of the Journal for a colour
version of this figure.}}}\label{fig:cm}   
\end{centering}
\end{figure}

Theoretical modeling does not strictly predict a universal
profile. Lower mass systems should have, on average, higher
concentration parameters (e.g., Navarro \etal~\cite{nfw97}; Bullock et
al.~\cite{bull01}), a consequence of their having formed
earlier. There also  appears to be a wide dispersion in the value of
$c$ in simulated haloes of a given mass, probably linked to scatter in
their formation epoch  (e.g. Wechsler \etal~\cite{wec02}).  To further
test the theory of the gravitational collapse, we now compare the
observed concentration parameters with the theoretical relation,
$\bar{c}(M)$, taking into account  the typical dispersion around it.  

Fig.~\ref{fig:cm} shows the concentration parameter $c_{200}$  of the
5 clusters under consideration plotted versus $\Mv$. We have also
added values from the literature for local clusters ($z \leq 0.15$)
with $c_{200}$ measured to a precision of better than $\pm 20$ per
cent.  The references used are given in the caption of the figure.
This sample, although still of modest size, spans a wide range of
cluster mass: from $M_{200} = 1.2 \times 10^{14} \hs^{-1} \msol$ to
$M_{200}  = 1.8 \times 10^{15} \hs^{-1} \msol$. 

Some care is required in comparing to results from numerical
simulations. The concentration parameter is commonly defined as  the
ratio of the `virial' radius to the scaling radius of the NFW model:
$c=r_{\rm vir}/\rs$, and the mass as the `virial' mass within $r_{\rm
vir}$. However different conventions for $r_{\rm vir}$  are used in
the literature: i) $r_{\rm vir} = r_{200}$  
where $\rv$ is the radius enclosing 200 times the {\it critical}
universe density (Navarro \etal~\cite{nfw97}); ii) $r_{\rm vir} =
r_{\Delta}$, the radius enclosing $\Delta$ times the critical density,
where $\Delta$ is derived from the spherical collapse model  (Eke
\etal~\cite{eke01}; Bullock et al.~\cite{bull01} ; Wechsler
\etal~\cite{wec02}; Zhao \etal~\cite{zha03}); iii) $r_{\rm vir} =
r_{200 \Omega}$ the radius enclosing 200 times the {\it mean} universe
density (Dolag \etal~\cite{dol04}).   

These different conventions lead to different concentrations and
masses for a $\Lambda$CDM Universe since $\rv <  r_{\Delta} < r_{200
\Omega}$ (see Huffenberger \& Seljak~\cite{hs03}). When the various
results are translated to a common virial radius convention, there is
good agreement on the $\bar{c}(M)$ relation (see e.g. Zhao
\etal~\cite{zha03} and Dolag \etal~\cite{dol04}).  We thus only
consider the recent work of Dolag \etal (\cite{dol04}), who found that
the mean concentration is well fitted  by a power law function of the
mass, $(1+z)\bar{c} = c_{0} (M/M_{0})^{\alpha}$ with $\alpha \sim
-0.1$ ($\Lambda$CDm cosmology, $\sigma_{8} = 0.9$). Note that these
authors used the convention $r_{\rm vir} = r_{200 \Omega}$ while we
use $r_{\rm vir} = r_{200}$.  We thus converted (using the NFW
profile) their $c_{200 \Omega}$ and $M_{200 \Omega}$ values to
$c_{200}$ and $M_{200}$ values.  At $z=0$ we found $c_{200}/c_{200
\Omega} \sim 0.6$ and $M_{200}/M_{200 \Omega} = 0.73 - 0.7$ in the
mass range considered here. The corresponding $\bar{c}_{200}(\Mv)$
relation is plotted as a full line in Fig.~\ref{fig:cm}.  Dolag \etal
(\cite{dol04}) also studied the scatter at a given mass: the
concentration  follows a log-normal distribution with a dispersion of
$0.2$, in good agreement with previous studies (Eke
\etal~\cite{eke01}; Bullock et al.~\cite{bull01}). The curves
corresponding to the mean plus/minus the standard deviation are
plotted as dotted lines in Fig.~\ref{fig:cm}.  The concentration
depends on redshift, so we have also plotted the $\bar{c}_{200}(\Mv)$
relation at $z=0.15$,  the highest redshift for our sample (dashed
line).  The evolution effect is negligible as compared to the
dispersion.  

It is clear from Fig.~\ref{fig:cm} that there is an excellent
agreement between the observed concentration parameters and the
theoretical predictions. Taking into account the measurement errors
and expected intrinsic scatter, there is no obvious deviation from the
theoretical $c(M)$ curve\footnote{In our published work on A1983
(Pratt \& Arnaud~\cite{pa03}) we mentioned that the concentration
parameter is lower than expected for a cluster of this mass. This is
not strictly true, as we in fact compared our $c_{200}$ value  with
the $c_{\Delta}$  value of Eke \etal (\cite{eke01}) for a $\Lambda$CDM
cosmology, and did not take into account  the expected theoretical
dispersion.}. However, the present data set does not allow us to check
the expected increase of concentration parameter with decreasing
mass. The mass dependence is small in the mass range under
consideration, being of the order of the typical dispersion (see
Fig.~\ref{fig:cm}). We could only test the expected $c(M)$ dependence
with a much larger sample (preferably with individual $c_{200}$ values
measured to a precision smaller than the scatter).  

\subsection{The gas specific physics}
\label{sec:disgas}

\subsubsection{A universal entropy profile?}

The present study confirms the remarkable self-similarity in the shape
of the entropy profiles down to low mass ($\kT \sim 2~ \keV$),
suggested by the study of Ponman \etal (\cite{psf03}) and our previous
\xmm\ test-case comparison of the cool cluster A1983 and the massive
cluster A1413 (Pratt \& Arnaud~\cite{pa03}).  Note that the accuracy
of the entropy profiles, which are well resolved  from $0.01~\rv$ up
to $0.5~\rv$, has allowed us to perform a direct check of the
self-similarity. In contrast, the {\it ROSAT/ASCA} study of Ponman
\etal (\cite{psf03}) had to rely both on extrapolations beyond the
detection radius (particularly at low mass) and stacking
analysis. These \xmm\ data have also allowed us to quantify the shape
of the entropy profile, which is found to be only slightly shallower
than expected in the pure shock heating model.  

Our study  also confirms that the $S$--$T$ scaling relation is
shallower than predicted by the purely gravitational model. The
dispersion in the scaled profiles is decreased when we use the
empirical relation $S\propto T^{0.65}$, estimated by Ponman \etal
(\cite{psf03}) at $0.1~\rv$.  We note that the normalisation of this
relation is consistent between the {\it ROSAT/ASCA} sample and the
present \xmm\ analysis. Ponman \etal (\cite{psf03}) obtained
$S(0.1\rv)\sim530~T_{10}^{0.65}~\hs^{-1/3} \keV~{\rm cm^{2}}$ (their
Figure 4). The normalisation is only  $13$ per cent higher than the
value,  we derive from the mean scaled entropy at that radius
($470~\hs^{-1/3} \keV~ {\rm cm^{2}}$; see Eq.~\ref{eq:sr}). This is
well within the typical dispersion.  Note that the self-similarity of
form found here implies that the slope of the $S$--$T$ relation (also
studied by Ponman \etal (\cite{psf03}) at $r_{500}$) should not depend
on  scaled radius, at least up to $0.5\rv$ (the maximum radius
considered here).  Its normalisation of course does:  from
Eq.~\ref{eq:sr}, the normalisation at $0.5~\rv$ is $\sim
2100~\hs^{-1/3}~\keV~{\rm cm^{2}}$. 

An alternative scaling based on preheating models was proposed by Dos
Santos \& Dor\'e~(\cite{dd}), in which the entropy should scale
according to $S \propto (1 + T/T_0)$, with $T_0~\sim 2~\keV$. Our
present sample does not allow us to distinguish between this scaling
and that of Ponman et al.~(\cite{psf03}). In fact, the difference
between scaling by $S \propto T^{0.65}$ and scaling by $S \propto (1 +
T/T_0)$ is only $\sim 10$ per cent in the 2 to 10 keV domain, which
explains why Ponman et al.~(\cite{psf03}) found good agreement when
their entropy profiles were scaled by the latter factor (their
Fig.~2). We further note that the Dos Santos \& Dor\'e scaling is
based on a preheating model where the factor $T_0$ was fixed to
recover the limiting entropy value proposed by Lloyd-Davies et
al.~(\cite{lpc}). We now know that such models are inconsistent with
both the observed power law $S$--$T$ behaviour and the lack of
constant entropy cores in cool systems. 

Our sample is still of modest size, and the dispersion analysis on the
scaled entropy profiles should not be overinterpreted.  It is a first
attempt to quantify the degree of self-similarity, as such better than
a simple comparison of the profiles by eye. While the analysis has
begun strongly to suggest self-similarity at low mass, our conclusion
on the self-similarity between poor and hot clusters still relies on
the assumption that A1413 is a fair representative of massive
clusters. The dispersion study does show that the $S\propto T^{0.65}$
scaling is better than the standard scaling, but does not allow us to
gain any further constraints on the exact slope of the $S$--$T$
relation (our present sample being too small, particularly with only
one cluster at high temperature).  We recall that Ponman \etal
(\cite{psf03}) give a slope of $\alpha = 0.65 \pm 0.05$.  

We feel it should also be pointed out that the degree of similarity
emphasised here does not imply a strict scaling of all the gas
properties.  

Firstly, there is large dispersion in the central region, $r \lesssim
0.05 \rv$, roughly corresponding to the cooling core region. For
instance  the entropy of A1991 and A1983 at $0.01~\rv$ differ by an
order of magnitude. These huge differences, also observed in \chandra
data by Sun \etal (\cite{sun04}), are likely to reflect the large
variety of cooling core histories, such as would be expected from the
complex interplay between cooling and central AGN activity.  

Secondly, relatively modest  differences in entropy translate into
much larger differences in other quantities, such as the X--ray
luminosity. This is because the luminosity is much more sensitive to
the gas density than the entropy:  $L_{\rm X} \propto \ne^{2}$ while
$S \propto \ne^{-2/3}$, so that   $L_{\rm X} \propto S^{-3}$ for
clusters of the same temperature and internal structure.  This has
important consequences for sources of the scatter in the $L_{\rm
X}$--$T$ relation,  even after the exclusion of the cooling core
region.   
To further illustrate this point, we can compare in greater detail the
3 poor clusters A1991, A1983 and MKW9, which  have {\it practically
identical temperatures\/} (see Tab.~\ref{tab:glob}).  
Above $0.1~\rv~(\sim 2\arcmin)$, safely outside the cooling core
region, the temperature profiles are isothermal
(Fig.~\ref{fig:tprofs}) and the difference in the entropy profiles
reflects differences in the gas density profiles.  MKW9 has a higher
entropy {\it throughout\/} its ICM, at least up to the detection
limit, than either A1991 or A2717, by a factor $\sim 1.4-1.7$
(Fig.\ref{fig:e065}). Note that this difference corresponds  to  $\pm
(1.-1.5) \sigma$ around the mean.   From the simple argument, $L_{\rm
X} \propto S^{-3}$, we expect  the luminosity of MKW9 to be
considerably lower (by a factor $3-5$)  than the luminosity of A1991
or A2717.  A direct computation of the luminosity of the three
clusters shows that this is indeed the case. The luminosities of A2717
and A1991 within the virial radius, excluding the central $0.1\rv$
region, differ by less than $20$ per cent, while the luminosity of
MKW9 is 4 times lower.

In spite of  the above caveats, the picture that emerges from the
present \xmm\ study, together with previous constraints on the slope
of the $S$--$T$ relation (Ponman et al.~\cite{psf03}), is that: 

\begin{itemize}
\item
There is no break of self-similarity in the entropy profiles down to $\kT \sim 2 \keV$ and in the radial range $0.05~\rv < r < 0.5~\rv$ 
\item 
In that radius and temperature domain, the entropy typically behaves  as $S(r) \propto T^{0.65\pm0.05} (r/\rv)^{0.94\pm0.14}$,  with a normalisation $S\sim500~\hs^{-1/3}~\keV~{\rm cm^{2}}$ at $T=10~\keV$ and $r=0.1~\rv$.
\item
The best guess $1\sigma$ dispersion in the scaled entropy profiles  is $\sim 25$ per cent. 
\end{itemize}

\subsubsection{Comparison with theoretical expectations}

The current view is that the departures of the gas properties from the
standard self-similar picture are due to non-gravitational
processes. However, in principle, these departures could also be due
to some flaw in the modeling of the gravitational collapse itself,
e.g.  the underlying dark matter component does not obey
self-similarity.  We would like first to emphasize that the present
study,  which provides the first precise estimates of mass profiles
for a significant number of low mass clusters, demonstrates that this
is not the case (see Sect.~\ref{sec:disdm}).  However, this does not
mean that the gravitational heating of the gas is perfectly
understood.  For instance, Valegeas \etal (\cite{val03}) suggested
that shock heating at very large scales might contribute significantly
to the cluster entropy. 

It is now fairly clear that pure cooling or simple pre-heating models
fail to explain the observed cluster properties. This has already been
greatly discussed in the literature (e.g. see Ponman
\etal~\cite{psf03}), and we will only briefly comment on it. 

Spherical pre-heating models predict  both a break in the $S-T$
relation and large isentropic cores (e.g Tozzi \& Norman~\cite{tn01},
Figures 5 and 17)  that are  simply not seen, as the present study
confirms.  This does not mean that pre-heating does not play a
role. As pointed out by Voit \etal (\cite{voit03}) and Ponman \etal
(\cite{psf03}), pre-heating may also affect the generation of
intracluster entropy by smoothing the accreted matter distribution, an
effect not taken into account in simple spherical models. This effect,
in combination with cooling, could explain cluster entropy properties
(Voit \& Ponman~\cite{vp03}), although cosmological simulations are
required to fully assess this explanation. 

Pure cooling models are not actually discrepant with the entropy
properties outlined above. They can explain the S-T relation at
$0.1\rv$ (Muawong \etal~\cite{mua02}, Fig.~4; Kay~\cite{kay04},
Fig.~1; see also Voit \etal~\cite{voit02}), and do not predict a
strong break of self-similarity of the entropy profiles (except
perhaps below $\kT\sim 1~\keV$, Dav\'e \etal~\cite{dav02}, Fig.5),
although they may slightly overpredict the entropy at very large radii
(Kay~\cite{kay04}).  The main problem with pure cooling models, as
shown by the authors, is that they tend to suffer from overcooling
(e.g., Muanwong \etal~\cite{mua02}, Dav\'e et al.~\cite{dav02}), which
is at odds with the observed mass fraction of the stellar component
(Balogh et al.~\cite{bal01}).  Furthermore, it seems plausible that
the feedback be associated with the cooling (Voit and
Bryan~\cite{vb01}).

Thus the current consensus is that some combination of cooling and
feedback acts to modify the entropy in clusters.  The present entropy
profile results can be compared with the predictions of the recent
large scale numerical simulations of Borgani \etal (\cite{borg04}) and
Kay (\cite{kay04}). Borgani et al.'s simulations incorporate a
physically motivated model of star formation and SNII driven galactic
winds. The heating was found to be insufficient, leading to a steeper
$S-T$ relation than observed.  Moreover, although the simulated
entropy profiles were very similar in shape, their slope is shallower
than found in our study: behaving roughly as $S(r) \propto r^{0.73}$
(from Borgani \etal \cite{borg04}, Figure 14). This is linked to
strongly decreasing temperature profiles, which, as noted by the
authors, is at odds with the observations.  Targeted feedback models
(Kay \etal~\cite{kay03}; Kay~\cite{kay04}) seem to be more successful
in reproducing the observed properties of cluster entropy profiles. In
this phenomenological model, a fraction of  particles which are about
to undergo cooling are reheated to a fixed entropy level. When this
level is fixed to about $ 1000~\keV~{\rm cm^2}$, this model reproduces
the $S-T$ relation at $0.1 \rv$ (and at $0.5 \rv$). We also note the
remarkable agreement between these simulations and our observations
when comparing the entropy profile slope: both are slightly shallower
than the shock-heating case. For instance the entropy profile of a $3
\keV$ cluster above $0.05\rv$  is essentially a power law, with an
entropy of $\sim 100 ~\keV~{\rm cm^2}$ and $1000 ~\keV~{\rm cm^2}$ at
$0.05 \rv$ and $0.5 \rv$ respectively. This can be compared to $\sim
120~\keV~{\rm cm^2}$ and $1000 ~\keV~{\rm cm^2}$ from our results
(Eq.~\ref{eq:sr}).  It remains of course to find a physical motivation
for this type of feedback.  


\section{Conclusion}

We have presented results from new \xmm\ observations of three poor
clusters (A1991, A2717 and MKW9) having similar temperatures ($\kT =
2.65$, 2.53 and 2.58 keV), and similar redshifts ($0.04 < z < 0.06$).
The paper concentrates on the properties of the mass and entropy
profiles, which we were able to map up to $\sim 0.5~\rv$.  We then
combined these data with previously published data on A1983 ($\kT =
2.2$ keV) and on the massive cluster A1413 ($\kT = 6.5$ keV), and
examined the scaling properties of the profiles. 
The emerging picture from these XMM observations is that local
clusters do form a self-similar population down to low mass ($\Mv \sim
10^{14}~\hs^{-1}~\msol$ or $\kT \sim 2~\keV$).  

Our study has provided clear evidence that the dark matter profile of
local clusters is nearly universal and presents a central cusp, as
predicted by numerical simulations.  The concentration parameter of
the clusters, and of other massive clusters from the literature,  were
found to be consistent with the $c_{200}$--$M_{200}$ relation derived
from numerical simulations  for a $\Lambda$CDM comology, in the mass
range $\Mv = [1.2 \times10^{14}-1.9 \times10^{15}]~h_{70}^{-1}\msol$
(taking into account the measurement errors and expected intrinsic
scatter). This excellent agreement with theoretical predictions shows
that the physics of the dark matter collapse is basically understood.

Except in the very centre, the entropy profiles of these clusters are
self-similar in shape, with close to power law behaviour in the
$0.05~\rv < r < 0.5~\rv$ range. The slope is slightly shallower than
predicted by shock heating models, $S(r)\propto r^{0.94\pm0.14}$. We
have confirmed that the $S$--$T$ relation is shallower than in the
purely gravitational model, and has a normalisation consistent with
that found from previous {\it ROSAT/ASCA} studies. The  entropy
scaling behaviour is summarized in our Eq.~\ref{eq:sr}. We emphasize
that the self-similarity of shape is a strong new constraint, and
simple pre-heating models can already be ruled out. In addition to the
gravitational effect, the gas history probably depends on the
interplay between cooling and various galaxy feedback mechanisms.  

The shape and scaling properties of the mass and entropy profiles of
clusters are the key observational constraints to be considered by
theoretical work. These profiles reflect respectively the physics of
the gravitational collapse and thermodynamic history of the gas.  They
can now be measured with \xmm\  in both massive and low mass clusters
over exceptionally wide radial  ranges.  Larger cluster samples are
still required to firmly establish the exact slopes and intrinsic
scatter of the scaling laws, the $S$--$T$, $\Mv$--$T$ and $c$--$M$
relations, and to confirm the self-similarity of form of the profiles
over the full mass range. 

\begin{acknowledgements}
We thank Doris Neumann for her contribution to the early part of this
project, and she, Etienne Pointecouteau and Klaus Dolag for useful
discussions. We thank the referee for comments. GWP acknowledges
support from the French Space Agency (CNES), and from the European
Commission through a Marie Curie Intra-European Fellowship under the
FP6 programme (contract no. MEIF-CT-2003-500915). The present work is
based on observations obtained with {\it XMM-Newton} an ESA science
mission with instruments and contributions directly funded by ESA
Member States and the USA (NASA). This research has made use of the
NASA's Astrophysics Data System Abstract Service; the SIMBAD database
operated at CDS, Strasbourg, France; the High Energy Astrophysics
Science Archive Research Center Online Service, provided by the
NASA/Goddard Space Flight Center and the Digitized Sky Surveys
produced at the Space Telescope Science Institute.

\end{acknowledgements}

\end{document}